\documentclass[aps,prd,superscriptaddress,A4paper,preprintnumbers,nofootinbib]{revtex4}
\usepackage{amssymb,amsmath,graphicx,epsfig,bm,color}
\usepackage{bm}
\renewcommand{\d}{\mathrm{d}}
\newcommand{\be}{\begin{equation}}
\newcommand{\ee}{\end{equation}}
\newcommand{\bea}{\begin{eqnarray}}
\newcommand{\eea}{\end{eqnarray}}

\newcommand{\bfk}{\mbox{\boldmath $k$}}

\newcommand{\bfp}{\mbox{\boldmath $p$}}

\newcommand{\pup}{p^\uparrow}

\def\lsim{\mathrel{\rlap{\lower4pt\hbox{\hskip1pt$\sim$}}\raise1pt\hbox{$<$}}}
\def\gsim{\mathrel{\rlap{\lower4pt\hbox{\hskip1pt$\sim$}}\raise1pt\hbox{$>$}}}
\def\nostrocostruttino#1\over#2{\mathrel{\mathop{\kern 0pt \rlap
{\hbox{$#1$}}} \hbox{\kern-.135em $#2$}}}

\begin{document}

\title{Unraveling the Gluon Sivers Function in Hadronic Collisions at RHIC}
\author{Umberto~D'Alesio}
\email{umberto.dalesio@ca.infn.it}
\affiliation{Dipartimento di Fisica, Universit\`a di Cagliari, Cittadella Universitaria, I-09042 Monserrato (CA), Italy}
\affiliation{INFN, Sezione di Cagliari, C.P.~170, I-09042 Monserrato (CA), Italy}
\author{Carlo~Flore}
\email{carlo.flore@ca.infn.it}
\affiliation{Dipartimento di Fisica, Universit\`a di Cagliari, Cittadella Universitaria, I-09042 Monserrato (CA), Italy}
\affiliation{INFN, Sezione di Cagliari, C.P.~170, I-09042 Monserrato (CA), Italy}
\author{Francesco~Murgia}
\email{francesco.murgia@ca.infn.it}
\affiliation{INFN, Sezione di Cagliari, C.P.~170, I-09042 Monserrato (CA), Italy}
\author{Cristian~Pisano}
\email{cristian.pisano@ca.infn.it}
\affiliation{Dipartimento di Fisica, Universit\`a di Cagliari, Cittadella Universitaria, I-09042 Monserrato (CA), Italy}
\affiliation{INFN, Sezione di Cagliari, C.P.~170, I-09042 Monserrato (CA), Italy}
\author{Pieter~Taels}
\thanks{Now at INFN, Sezione di Cagliari}
\email{pieter.taels@ca.infn.it}
\affiliation{INFN, Sezione di Pavia, Via Bassi 6, I-27100 Pavia, Italy}

\date{\today}

\begin{abstract}
We study the transverse single-spin asymmetries for $p^\uparrow  p\to \pi\, X$ and $p^\uparrow  p\to \gamma\, X$ within the so-called color gauge invariant generalized parton model (CGI-GPM) which, in addition to spin and transverse momentum effects, includes initial and final state interactions with the polarized proton remnants. We compute all relevant contributions, focusing in particular on the process dependence of the gluon Sivers function, which, for these processes, can always be expressed as a linear combination of two independent, universal terms. This study extends and completes a previous one, where only quark initiated partonic processes were considered.
We then perform a combined phenomenological analysis of RHIC data on transverse single-spin asymmetries in $p^\uparrow  p\to \pi\, X$ and $p^\uparrow p\to D\, X$, putting the first preliminary constraints on these two gluon Sivers functions. We show how their size can be estimated by means of these data, and use our results to provide predictions for the process $p^\uparrow p\to J/\psi\,X$, comparing them with data, and $p^\uparrow  p\to \gamma\, X$, for which experimental information will soon become available. Corresponding estimates within the simpler GPM approach, without initial and final state interactions and with a single universal gluon Sivers function, are also given, showing that a clear discrimination between these two models is, for the moment, not possible.
\end{abstract}

\maketitle

\section{\label{1}Introduction}

Among the various transverse momentum dependent parton distribution and fragmentation functions (TMDs for short), the Sivers function~\cite{Sivers:1989cc,Sivers:1990fh} is of great interest, both experimentally and theoretically. It is related to the asymmetry in the azimuthal distribution of unpolarized quarks and gluons inside a high-energy proton that is transversely polarized with respect to its momentum. As such, it can in turn give rise to azimuthal asymmetries of the produced particles in high-energy scattering processes initiated by transversely polarized protons. Moreover, the Sivers function is known to be very sensitive to the color exchanges among initial and  final states, and  to the color flow in the scattering processes. These peculiar properties have a clear signature~\cite{Collins:2002kn,Brodsky:2002rv}, providing  a strong test of the TMD formalism.

A first evidence of a nonzero Sivers distribution for quarks has come from data on single spin asymmetries for semi-inclusive deep inelastic processes (SIDIS), measured by the HERMES Collaboration at DESY~\cite{Airapetian:2004tw}, and confirmed later by the COMPASS Collaboration at CERN~\cite{Adolph:2012sp}. Nowadays, thanks to a continuous and dedicated experimental investigation and to new phenomenological extractions, it can be considered established.

The knowledge of the quark Sivers function, quite important by itself, provides an indirect constraint on the much less known gluon Sivers function by means of the Burkardt sum rule~\cite{Burkardt:2004ur}, which states that the transverse momenta of all unpolarized partons inside a transversely polarized proton add up to zero. Available parameterizations for the quark Sivers function~\cite{Anselmino:2005ea,Anselmino:2008sga} almost fulfill, within uncertainties, the Burkardt sum rule, pointing towards a small gluon contribution. This is consistent with theoretical arguments valid in the large-$N_c$ limit of QCD~\cite{Efremov:2004tp,Boer:2015vso}, according to which the gluon Sivers function should be suppressed by a factor $1/N_c$ with respect to the valence quark Sivers distributions at values of the light-cone momentum fraction $x$ of the order of $ 1/N_c$.

Turning now to the discussion of direct probes of the gluon Sivers effect, we note that a first extraction of the gluon Sivers function from very precise data on single spin asymmetries in $p^\uparrow p \to \pi^0\,X $ at central rapidities~\cite{Adare:2013ekj} has  been attempted in the framework of the generalized parton model (GPM)~\cite{DAlesio:2015fwo}. In this approach, the TMD formalism is applied even to single-scale processes and transverse momentum dependent distribution and fragmentation functions are conditionally taken to be universal. Although lacking a formal proof, the GPM is phenomenologically very successful in describing many processes for which data are available, see Refs.~\cite{Anselmino:1994tv, D'Alesio:2004up,Anselmino:2005sh,DAlesio:2007bjf, Anselmino:2013rya, Aschenauer:2015ndk,DAlesio:2017nrd}.

In the meantime, a Color Gauge Invariant formulation of the GPM, named CGI-GPM~\cite{Gamberg:2010tj,DAlesio:2011kkm,DAlesio:2013cfy} has been proposed, in which the effects of initial (ISI) and final (FSI) state interactions on the quark Sivers function are taken into account, within a one-gluon exchange approximation. As a result, the Sivers function for quarks becomes nonuniversal, and its process dependence can be absorbed into the partonic cross sections. Hence, in the calculation of physical observables, for example in proton-proton collisions, one can still use the quark Sivers functions obtained from SIDIS data, but they need to be convoluted with the {\em modified} partonic cross sections calculated in Ref.~\cite{Gamberg:2010tj}. In particular, the CGI-GPM can reproduce the expected opposite relative sign of the quark Sivers functions in SIDIS and in the Drell-Yan processes~\cite{Collins:2002kn,Brodsky:2002rv}.

In Ref.~\cite{DAlesio:2017rzj}, the CGI-GPM has been for the first time extended to the gluon Sivers function in the study of inclusive $J/\psi$ and $D$ meson production in proton-proton collisions at RHIC. These processes, as compared to pion production, have the advantage of probing gluon TMDs directly, since quark induced subprocesses can be safely neglected in the kinematical regions considered. Similarly to the quark case, the process dependence of the gluon Sivers function can still be absorbed into the hard partonic cross sections. However, one needs to introduce two different classes of modified partonic cross sections, corresponding to the two different ways in which a color-singlet state can be formed out of three gluons, {\it i.e.}\ either through an antisymmetric or a symmetric color combination. Each one of them has to be convoluted with a different gluon Sivers distribution. These two universal and independent distributions are named, respectively, the $f$-type and $d$-type gluon Sivers functions~\cite{Bomhof:2006ra}, or $A_1$ and $A_2$ in the notation of Ref.~\cite{Buffing:2013kca}: the former is even under charge conjugation, while the latter is odd. It turns out that only the $f$-type distribution contributes to $J/\psi$ production, at least in the analyzed kinematical region where the color-singlet mechanism is dominant, while for $D$-meson production the $d$-type is the most relevant one~\cite{DAlesio:2017rzj}. Corresponding studies, within the GPM framework only, have been presented in Ref.~\cite{Anselmino:2004nk} and later on in Refs.~\cite{Godbole:2016tvq,Godbole:2017syo}.

In the present paper we extend the formalism of the CGI-GPM to the processes $p^\uparrow p \to \pi\, X$ and $p^\uparrow p \to \gamma\, X$. We calculate all modified partonic cross sections induced by gluons, needed for a re-analysis of the RHIC pion data of Ref.~\cite{Adare:2013ekj}. These results are therefore complementary to the quark-induced ones published in Ref.~\cite{Gamberg:2010tj}. Moreover, we perform a detailed phenomenological analysis and show how it is possible to disentangle and give an estimate of the size of the two gluon Sivers functions. To this end we study, in the same framework (see Ref.~\cite{DAlesio:2017rzj}), also the latest available data on inclusive $D$-meson production~\cite{Aidala:2017pum}. We then compare our new predictions for single spin asymmetries in $p^\uparrow p\to J/\psi\,X$ with the most recent RHIC data~\cite{Aidala:2018gmp} and give the corresponding theoretical estimates for the kinematics reachable at LHC with a fixed polarized target. Finally, we give predictions for the process $p^\uparrow p \to \gamma\, X$ currently under investigation at RHIC, for which data are expected in the near future.

The paper is organized as follows: in Section~\ref{theory} we present the leading order  partonic cross sections, within the framework of the CGI-GPM, for the gluon induced subprocesses that contribute to the Sivers asymmetry in $p^\uparrow p\to h\, X$ (Sect.~\ref{pion}) and in $p^\uparrow p \to \gamma\, X$ (Sect.~\ref{photon}). In Section~\ref{pheno} we perform a phenomenological analysis of available data on single spin asymmetries in $p^\uparrow p\to \pi\, X$ and $p^\uparrow p\to D\, X$ putting some reliable constraints on the gluon Sivers function, then in Section~\ref{predict} we present our predictions for the same observable in $p^\uparrow p\to J/\psi\, X$ (for which a comparison with data is possible) and $p^\uparrow p\to \gamma\, X$. Conclusions and final remarks are collected in Section~\ref{concl}.
The color factors needed for the calculation of the hard functions $H^{\text{Inc}}_{ab\to c d}$ within the CGI-GPM are listed in the two Appendices.

\section{Theoretical Framework}\
\label{theory}
The single-spin asymmetries (SSAs)  for the processes $p^\uparrow p \to h\, X$ and $p^\uparrow p \to \gamma\,  X$ are defined as follows
\begin{equation}
A_N \equiv \frac{\d\sigma^\uparrow-\d\sigma^\downarrow}{\d\sigma^\uparrow+\d\sigma^\downarrow} \equiv\, \frac{ \d\Delta\sigma}{ 2 \d\sigma}\,,
\end{equation}
where $\d\sigma^{\uparrow (\downarrow)}$ denotes the single-polarized cross section, in which one of the protons in the initial state is polarized along the transverse direction $\uparrow$ ($\downarrow$) with respect to the production plane.
As extensively studied in Ref.~\cite{Anselmino:2005sh}, within a TMD approach, the numerator of the asymmetry is mainly driven by only two contributions: the Sivers~\cite{Sivers:1989cc,Sivers:1990fh} and the Collins~\cite{Collins:1992kk} effects. Furthermore, in suitable kinematical regions, as we are going to discuss below, only the Sivers effect can be sizeable. Hence, the numerator of the asymmetry is sensitive to the quantity~\cite{Bacchetta:2004jz}
\bea
\Delta \hat f_{a/\pup}\,(x_a, \bfk_{\perp a}) &\equiv&
\hat f_{a/\pup}\,(x_a, \bfk_{\perp a}) - \hat f_{a/p^\downarrow}\,
(x_a, \bfk_{\perp a})\nonumber \\
\label{defsiv}
&=& \Delta^N f_{a/\pup}\,(x_a, k_{\perp a}) \> \cos\phi_a\nonumber \\
&=&  -2 \, \frac{k_{\perp a}}{M_p} \, f_{1T}^{\perp a} (x_a, k_{\perp a}) \>
\cos\phi_a \, ,
\eea
with $\hat f_{a/\pup}\,(x_a, \bfk_{\perp a})$ being the number density of partons $a$ with light-cone momentum fraction $x_a$ and transverse momentum $\bm k_{\perp a} = k_{\perp a} (\cos\phi_a, \sin\phi_a)$ inside the transversely polarized proton with mass $M_p$, which is taken to move along the $\hat z$-axis.  The Sivers distribution of parton $a$ is represented either by $\Delta^N f_{a/\pup}(x_a, k_{\perp a})$ or $f_{1T}^{\perp a} (x_a, k_{\perp a})$ and fulfills the following positivity bound
\begin{equation}
\vert \Delta^N f_{a/\pup}\,(x_a, k_{\perp a}) \vert  \le 2\,f_{a/p}\,(x_a, k_{\perp a})\,,~~{\rm or}~~
\frac{k_{\perp a}}{M_p}\, \vert f_{1T}^{\perp a} (x_a, k_{\perp a})\vert \le  f_{a/p}\,(x_a, k_{\perp a})~.
\label{eq:posbound}
\end{equation}
We note that, since $a$ can be either a quark (antiquark) or a gluon, the Sivers contribution to the asymmetry can be expressed as a sum of two terms, namely
\be
A_N = A_N^{\rm quark} + A_N^{\rm gluon}\,,
\ee
where quark (gluon) refers to the parton inside the polarized proton in the numerator of $A_N$.
The quark and gluon contributions to $A_N$ cannot be directly disentangled either in $p^\uparrow p \to \pi\, X$ or in $p^\uparrow p \to \gamma\, X$. For this reason, in our numerical studies, focused on the extraction of the gluon Sivers function, we will use all the available information on the quark Sivers functions coming from the analysis of azimuthal asymmetries in SIDIS processes.

In the next two subsections, we provide the explicit expressions of the numerators of the asymmetries for  $p^\uparrow p \to \pi\, X$ and $p^\uparrow p \to \gamma\, X$, respectively, in the CGI-GPM approach. The corresponding formulae for  $p^\uparrow p \to J/\psi\, X$ and $p^\uparrow p \to D\, X$ are given in Ref.~\cite{DAlesio:2017rzj}, where it was found that, for such processes, the gluon contribution to the asymmetry is dominant.

\subsection{SSA in $p^\uparrow p\to \pi\, X $ }
\label{pion}
Within the framework of the CGI-GPM, the numerator of the asymmetry is given by
\begin{align}
 \d\Delta\sigma^{\rm CGI-GPM}\, \equiv  &\,  \frac{E_\pi \, \d\sigma^\uparrow}{\d^3\bfp_\pi} -
\frac{E_\pi \, \d\sigma^\downarrow}{\d^3\bfp_\pi} \simeq  \frac{2\alpha_s^2}{s}\sum_{a,b,c,d} \int \frac{\d x_a \, \d x_b \, \d z}{ x_a \, x_b \, z^2 } \; \d^2\bfk_{\perp a} \, \d^2\bfk_{\perp b} \,
\,\d^3\bfk_{\perp \pi} \, \delta(\bfk_{\perp \pi}\cdot \hat{\bfp}_c) \, J(k_{\perp \pi})
\nonumber \\
 &\times  \left ( -\frac{k_{\perp a}}{M_p} \right  ) f^{\perp a}_{1 T}(x_a, k_{\perp a})\cos\phi_a
\> f_{b/p}(x_b, k_{\perp b})\,H^{\rm Inc}_{ab \to cd}
(x_a,x_b, \hat s, \hat t, \hat u) \> \delta(\hat s + \hat t + \hat u) \>  D_{\pi/c}(z, k_{\perp \pi}) \>,
\label{sivgen}
\end{align}
where $J(k_{\perp \pi})$ is a kinematical factor~\cite{D'Alesio:2004up} and $\hat s$, $\hat t $, $\hat u$ are the usual Mandelstam variables for the partonic subprocess $ab\to cd$. Furthermore,  $f_{b/p}(x_b, k_{\perp b})$ is the TMD distribution for an unpolarized parton $b$ inside the unpolarized proton, while $D_{\pi/c}(z, k_{\perp \pi})$ is the the fragmentation function of an unpolarized parton $c$ into a pion.
Finally, $H^{\rm Inc}_{ab \to cd}$ are the perturbatively calculable hard scattering functions. In particular, the ones for which $a$ is a quark or an antiquark, are well-known and can be found in Ref.~\cite{Gamberg:2010tj}, while the remaining ones have been evaluated here for the first time along the lines of Ref.~\cite{DAlesio:2017rzj}.
As already pointed out, in the CGI-GPM approach there are two independent gluon Sivers contributions: the $f$- and $d$-type.
The leading order (LO) explicit expressions for the hard functions corresponding to the gluon Sivers distribution $f_{1T}^{\perp g\, (f)}$ read
\begin{align}
&H_{gq\to gq}^{\text{Inc} \,(f)} = H_{g\bar{q}\to g\bar{q}}^{\text{Inc} \,(f)}  = -\frac{\hat s^2+\hat u^2}{4 \hat s \hat u}\, \bigg (  \frac{\hat s^2}{\hat t^2} + \frac{1}{N_c^2} \bigg ) \,,\\
& H_{gq\to qg}^{\text{Inc} \,(f)} = H_{g\bar{q}\to \bar{q} g}^{\text{Inc} \,(f)}  = -\frac{\hat s^4-\hat t^4}{4 \hat s \hat t\hat u^2} \,,      \\
&H_{gg\to q \bar q}^{\text{Inc} \,(f)}  = H_{gg\to \bar{q} q}^{\text{Inc} \,(f)}=-\frac{N_c}{4 (N_c^2-1)}\, \frac{\hat t ^2+ \hat u^2}{ \hat t \hat u} \,\bigg (\frac{\hat t^2}{\hat s^2}\, +\frac{1}{ N_c^2} \bigg )\,, \\
&H_{gg\to gg}^{\text{Inc} \,(f)} = \frac{N_c^2}{N_c^2-1}\,\bigg (\frac{\hat t}{\hat u} - \frac{\hat s}{\hat u}\bigg)\, \frac{(\hat s^2 +  \hat s \hat t + \hat t^2)^2}{\hat s^2 \hat t^2}\, ,
\end{align}
where $N_c$ is the number of colors. For the other gluon Sivers function $ f_{1T}^{\perp g\, (d)}$, one has
\begin{align}
&H_{gq\to gq}^{\text{Inc} \,(d)} = -H_{g\bar{q}\to g\bar{q}}^{\text{Inc} \,(d)}= \frac{\hat s^2 + \hat u^2}{ 4 \hat s \hat u} \, \bigg ( \frac{ \hat s^2-2\hat u^2}{\hat t^2} + \frac{1}{N_c^2}  \bigg ) \,,\label{eq:Hgqgq}\\
& H_{gq\to q g}^{\text{Inc} \,(d)} = -H_{g\bar{q}\to \bar{q} g}^{\text{Inc} \,(d)} = -\frac{\hat s^2 + \hat t^2}{4 \hat s \hat t } \, \left (\frac{\hat s^2 + \hat t^2}{\hat u^2}  - \frac{2}{N_c^2}    \right )  \,,\label{eq:Hgqqg}\\
& H_{gg\to q \bar q}^{\text{Inc} \,(d)}  = -H_{gg\to \bar{q} q}^{\text{Inc} \,(d)} =-\frac{N_c}{4 (N_c^2-1)}\, \frac{\hat t ^2+ \hat u^2}{ \hat t \hat u} \,\bigg ( \frac{\hat t^2-2 \hat u^2}{\hat s^2} + \frac{1}{N_c^2} \bigg )\,,\\
& H_{gg\to gg}^{\text{Inc} \,(d)} = 0\,\label{eq:Hgggg}.
\end{align}
More details on their calculation are given in Appendix~\ref{sec:cf-pion}. For comparison, we show the corresponding, well-known unpolarized hard functions,
\begin{align}
& H^U_{gq\to gq} =  - \frac{\hat s^2 + \hat u^2}{2 \hat s \hat u} \, \left ( \frac{\hat s^2 + \hat u^2}{\hat t^2}-\frac{1}{N_c^2} \right ) \,,\\
& H^U_{gg\to q \bar q} = \frac{N_c}{N_c^2-1}\, \frac{\hat t^2 + \hat u^2}{2 \hat t \hat u} \left ( \frac{\hat t^2 + \hat u^2 }{\hat s^2}  - \frac{1}{N_c^2} \right ) \,, \\
& H^U_{gg\to g g} = \frac{N_c^2}{N_c^2-1}\, \frac{(\hat s^4 + \hat t^4 + \hat u^4)(\hat s^2 + \hat t^2 + \hat u^2)}{\hat s^2 \hat t^2 \hat u^2}\,,
\end{align}
defined in such a way that
\begin{equation}
\frac{\d\hat\sigma}{\d\hat t} = \frac{ \pi \alpha_s^2}{\hat s^2}\, H^U_{ab\to c d}\, ,
\end{equation}
which appear in the denominators of the asymmetries.

\subsection{SSA in $p^\uparrow p \to \gamma\, X$}
\label{photon}
The numerator of the SSA for the process $p^\uparrow p \to \gamma\, X$ reads
\bea \!\!\!\!\! && \frac{E_\gamma \, \d\sigma^\uparrow}{\d^3\bfp_\gamma} -
\frac{E_\gamma \, \d\sigma^\downarrow}{\d^3\bfp_\gamma} \simeq  \frac{2\alpha\alpha_s e^2_q}{s}\sum_{a,b,d} \int \frac{\d x_a \, \d x_b }{ x_a \, x_b  } \; \d^2\bfk_{\perp a} \, \d^2\bfk_{\perp b}
\nonumber \\ \!\!\!\!\! &\times& \left ( -\frac{k_{\perp a}}{M_p} \right  ) f^{\perp a}_{1 T}(x_a, k_{\perp a})\cos\phi_a
\> f_{b/p}(x_b, k_{\perp b})\,H^{\rm Inc}_{ab \to \gamma d}
(x_a,x_b, \hat s, \hat t, \hat u) \> \delta(\hat s + \hat t + \hat u) \, .
\label{sivgeng}
\eea
As for $p^\uparrow p \to \pi\, X$, the partonic hard functions in which the parton $a$ inside the polarized proton is a quark or an antiquark are given in Ref.~\cite{Gamberg:2010tj}. For the gluon induced subprocesses, we find
\begin{align}
& H^{\text{Inc}\,(f)}_{gq\to \gamma q} =  H^{\text{Inc}\,(f)}_{g\bar{q}\to \gamma \bar{q}} = -\frac{1}{2} \,H^U_{gq\to \gamma q} 
\, , \label{eq:Hgqgammaq1}\\
& H^{\text{Inc}\,(d)}_{gq\to \gamma q} = -H^{\text{Inc}\,(d)}_{g\bar{q}\to \gamma \bar{q}} = \frac{1}{2} \,H^U_{gq\to \gamma q} 
\,,\label{eq:Hgqgammaq2}
\end{align}
for the $f$- and $d$-type gluon Sivers functions, respectively. The unpolarized hard function is given by
\begin{align}
H^U_{gq\to \gamma q} & =  H^U_{g\bar{q}\to \gamma \bar{q}} = \frac{1}{N_c} \left (-\frac{\hat u}{\hat s}  -\frac{\hat s}{\hat u} \right )\,,
\end{align}
and  is normalized such that the corresponding partonic cross section has the following form:
\begin{equation}
\frac{\d\hat\sigma}{\d\hat t} = \frac{ \pi \alpha\alpha_se^2_q}{\hat s^2}\, H^U_{g q\to \gamma q}\,.
\end{equation}

We refer to Appendix~\ref{sec:cf-gamma} for further details of the calculation.

\section{Phenomenology}
\label{pheno}

We are now able to devise a possible strategy to put the first reliable constraints on the two independent gluon Sivers functions within the CGI-GPM approach. To this aim, in Section~\ref{pheno-const} we will present a detailed analysis of SSA data in $p^\uparrow p\to \pi\, X$ and $p^\uparrow p\to D\, X$.
We will compare our findings with the available data, as well as with the corresponding results in the GPM scheme, as obtained in Ref.~\cite{DAlesio:2015fwo}. Finally, in Section~\ref{predict} we will show new predictions for SSAs in $p^\uparrow p\to J/\psi \,X$ and $p^\uparrow p\to \gamma\, X$.

\subsection{Constraints on the gluon Sivers functions from available data}
\label{pheno-const}

As discussed in the previous Section, in the CGI-GPM framework there are two universal and independent gluon Sivers functions (GSFs), the $f$- and $d$-type, and the phenomenological analysis appears more difficult with respect to the one in the GPM scheme. The reason is that, in principle, different combinations of these two contributions could lead to similar results and describe equally well the same set of data.
Therefore, in order to carry out this analysis we will have to use at least two independent sets of data. In particular, we will use the extremely precise and accurate data on SSAs in $pp$ collisions for inclusive pion production at mid-rapidity~\cite{Adare:2013ekj} and those for $D$-meson production~\cite{Aidala:2017pum} by the PHENIX Collaboration. They also collected SSA data for $J/\psi$ production~\cite{Aidala:2018gmp}, which we will compare against our estimates. From the phenomenological point of view, it is worth noticing that for the latter process, in the CGI-GPM approach, only the $f$-type contribution appears. Therefore, as it will become more clear in the following, it is important to consider additional processes, where also the $d$-type GSF plays a role.

All these processes have a common feature: the gluon initiated subprocesses dominate over the quark ones. As was already pointed out in Refs.~\cite{Anselmino:2006yq, DAlesio:2015fwo}, the SSA for inclusive pion production in $pp$ collisions at mid-rapidity is directly sensitive to the gluon Sivers distribution. In fact, the contribution involving the quark Sivers functions, as extracted from SIDIS azimuthal asymmetry data, is totally negligible -- this is true also in the CGI-GPM approach, as we will show in the following -- and all other effects, like the one driven by the Collins function, are washed out by integrations over the azimuthal phases.
Concerning the SSAs in $D$-meson production, as discussed in Ref.~\cite{DAlesio:2017rzj}, one has a clear and direct access to the GSF, due to the dominance of the $gg\to c\bar c$ channel.

Within our strategy, the first issue we address is to which extent the $f$- and $d$-type contributions are effectively relevant in the process under consideration. More precisely, we start with the observation that the numerators of the SSAs, Eqs.~(\ref{sivgen}) and (\ref{sivgeng}), contain three fundamental quantities: the azimuthal factor of the gluon Sivers function, $\cos\phi_a$ (with $\phi_a$ to be integrated over), the perturbatively calculable hard partonic parts, $H_{ab\to cd}$, and the unknown GSF, $f_{1T}^{\perp g}$. In order to explore the role played by the first two factors, we calculate the SSAs by maximizing the corresponding GSFs.
To do this we adopt the well-known Gaussian-like and factorized parametrization for the GSF, as follows:
\begin{equation}
\Delta^N\! f_{g/p^\uparrow}(x,k_\perp) =   \left (-2\frac{k_\perp}{M_p}  \right )f_{1T}^{\perp\,g} (x,k_\perp)  = 2 \, {\cal N}_g(x)\,f_{g/p}(x)\,
h(k_\perp)\,\frac{e^{-k_\perp^2/\langle k_\perp^2 \rangle}}
{\pi \langle k_\perp^2 \rangle}\,,
\label{eq:siv-par-1}
\end{equation}
where $f_{g/p}(x)$ is the standard unpolarized collinear gluon distribution,
\begin{equation}
{\cal N}_g(x) = N_g x^{\alpha}(1-x)^{\beta}\,
\frac{(\alpha+\beta)^{(\alpha+\beta)}}
{\alpha^{\alpha}\beta^{\beta}}\,,
\label{eq:nq-coll}
\end{equation}
with $|N_g|\leq 1$, and
\begin{equation}
h(k_\perp) = \sqrt{2e}\,\frac{k_\perp}{M'}\,e^{-k_\perp^2/M'^2}\,.
\label{eq:h-siv}
\end{equation}

Alternatively, if we define the parameter
\begin{equation}
\rho = \frac{M'^2}{\langle k_\perp^2 \rangle +M'^2}\, ,
\label{eq:rho}
\end{equation}
such that $0< \rho < 1$, then  Eq.~(\ref{eq:siv-par-1}) becomes
\begin{equation}
\Delta^N\! f_{g/p^\uparrow}(x,k_\perp) =   2 \,  \frac{\sqrt{2e}}{\pi}   \, {\cal N}_g(x)\,f_{g/p}(x)\,\sqrt{\frac{1-\rho}{\rho}}\,k_\perp\,
\frac{e^{-k_\perp^2/ \rho\langle k_\perp^2 \rangle}}
{\langle k_\perp^2 \rangle^{3/2}}~.
\label{eq:siv-par}
\end{equation}

With these choices, assuming that the unpolarized TMD gluon distribution is given by
\be
f_{g/p}(x,k_\perp) = f_{g/p}(x) \frac{e^{-k_\perp^2/\langle k_\perp^2 \rangle}}{\pi \langle k_\perp^2 \rangle}\,,
\ee
the Sivers function automatically fulfills its proper positivity bound for any $(x,k_\perp)$ values (see Eq.~(\ref{eq:posbound})). Analogously, for the unpolarized TMD fragmentation function (for a parton $c$) we use~\cite{Anselmino:2005nn}
\be
D_{\pi/c}(z,k_{\perp\pi}) = D_{\pi/c}(z)\,
\frac{e^{-k_{\perp\pi}^2/\langle k_{\perp\pi}^2 \rangle}}
{\pi \langle k_{\perp\pi}^2 \rangle}
\quad\quad\quad \langle k_{\perp\pi}^2\rangle = 0.20\, {\rm GeV}^2 \>.
\ee
In this analysis we adopt the CTEQ6-LO parametrization~\cite{Pumplin:2002vw} for the unpolarized gluon distribution, $f_{g/p}(x)$, with the factorization scale equal to the pion transverse momentum, $p_T$, and the leading-order DSS set for the collinear fragmentation functions~\cite{deFlorian:2007aj}. Notice that all TMDs defined above evolve with the hard scale through the scale dependence of the collinear distributions entering in their parameterizations, that is following a DGLAP evolution.

The first $k_\perp$-moment of the Sivers function is also of relevance:
\be
\Delta^N \! f_{g/p^\uparrow}^{(1)}(x) = \int \d^2 \bm{k}_\perp \frac{k_\perp}{4 M_p} \Delta^N \! f_{g/\pup}(x,k_\perp) \equiv - f_{1T}^{\perp (1) g}(x) \, .
\label{siversm1}
\ee
Adopting the parameterization of Eqs.~(\ref{eq:siv-par-1})-(\ref{eq:h-siv}),
\be
\Delta^N \! f_{g/p^\uparrow}^{(1)}(x) = \frac{\sqrt{\frac{e}{2}} \ \langle k_\perp^2 \rangle M'^3}{M_p (\langle k_\perp^2 \rangle + M'^2)^2}  \ {\cal N}_g(x)  f_{g/p}(x) =
\sqrt{\frac{e}{2}} \frac{\sqrt{\langle k_\perp^2\rangle}}{M_p} \sqrt{\rho^3(1-\rho)}\ {\cal N}_g(x)  f_{g/p}(x)
\, .
\label{siversm2}
\ee

In Ref.~\cite{DAlesio:2015fwo} a single value $\langle k_\perp^2\rangle = 0.25$ GeV$^2$~\cite{Anselmino:2005nn} was adopted, the same for the unpolarized quark and gluon TMDs, while the parameters $N_g$, $\alpha$, $\beta$, $\rho$ were fitted to the data, within the GPM scheme. Here, following Ref.~\cite{DAlesio:2017rzj}, for the unpolarized gluon TMD we use a different value, $\langle k_\perp^2\rangle = 1$ GeV$^2$. This, indeed, gives a better account of the unpolarized cross sections for $J/\psi$ production at not so large $p_T$ values, still allowing a good description, for instance, of the inclusive pion production. For this reason, we have reanalysed the same set of data within the GPM approach, getting results very similar to those reported in Ref.~\cite{DAlesio:2015fwo}, although with slightly different parameters:
\begin{equation}
\label{eq:par_gsf_GPM}
N_g=0.25\,, \hspace*{1cm} \alpha = 0.6\,, \hspace*{1cm}\beta = 0.6\,, \hspace*{1cm}\rho = 0.1  \,.
\end{equation}

Notice that an equally good description of pion SSA data can be obtained even with different sets of the above parameters, that are strongly correlated among each other. While this could imply very different $k_\perp$ dependences of the GSF, its first $k_\perp$-moment remains almost unchanged in the range of $x$ probed by data ($10^{-3}\le x \le 0.4$).

Moving to the CGI-GPM approach, in order to maximize the effects of the GSFs, we \emph{saturate} the positivity bound for their $x$-dependent parts (i.e.~we take ${\cal N}_g(x)=\pm 1$) and adopt the value $\rho = 2/3$~\cite{DAlesio:2010sag} in Eq.~(\ref{eq:siv-par}). 

For the $x$-dependent part of the GSF one can also use the following notation
\begin{equation}
\Delta^N\! f_{g/p^\uparrow}(x) =   2\, {\cal N}_g(x)\,f_{g/p}(x) \,,
\label{eq:siv-par-x}
\end{equation}
which, for ${\cal N}_g(x)=\pm 1$, implies $\Delta^N\! f_{g/p^\uparrow}(x) = \pm 2 f_{g/p}(x)$.

\begin{figure}[t]
\begin{center}
\includegraphics[trim = 1.cm 0cm 1cm 0cm, width=8.5cm]{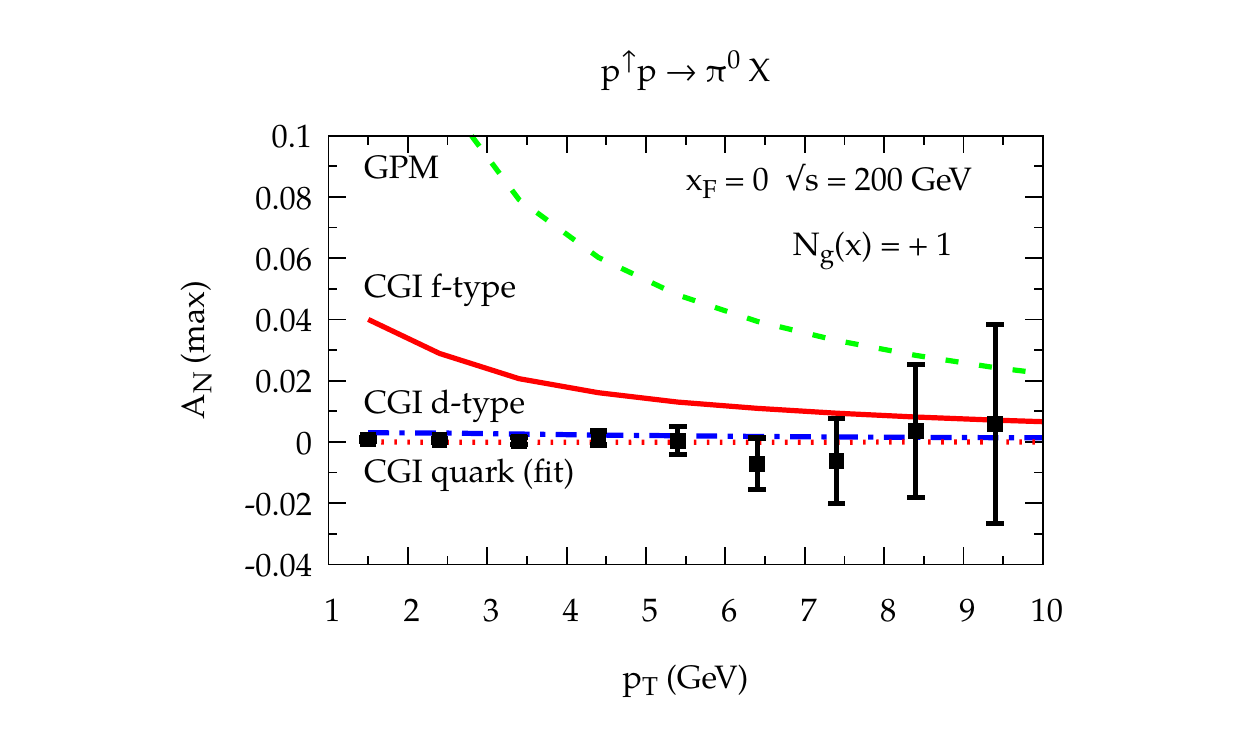}
\includegraphics[trim = 1.cm 0cm 1cm 0cm,width=8.5cm]{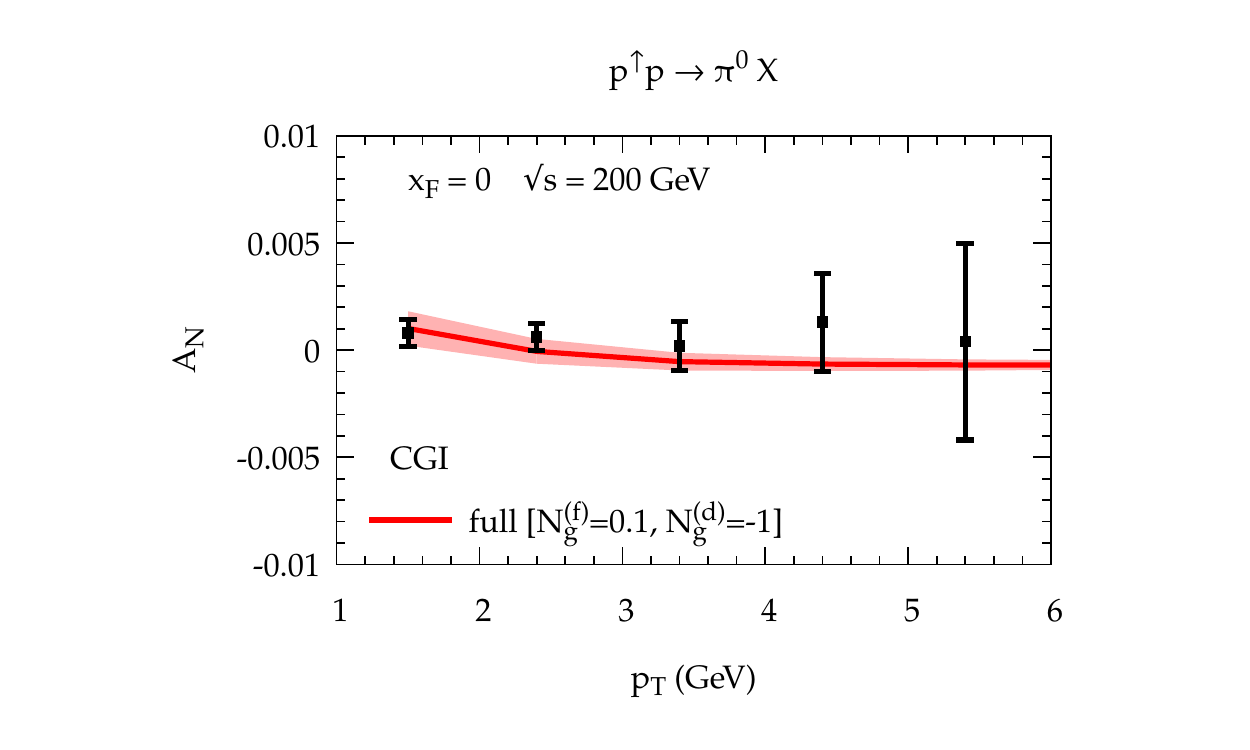}
\caption{Left panel: maximized gluon Sivers contributions (${\cal N}_g(x) =+1$) to $A_N$ for the process $p^\uparrow p\to \pi^0\, X$ at $\sqrt s=200$ GeV and mid-rapidity as a function of $p_T$ within the GPM (green dashed line) and the CGI-GPM approaches: $f$-type (red solid line) and $d$-type (blue dot-dashed line). The quark Sivers contribution within the CGI-GPM scheme, as extracted from SIDIS data, is also shown (red dotted line). Right panel: $A_N$ estimates, in the moderate $p_T$ range, obtained adopting a suitably reduced $f$-type GSF ((${\cal N}_g^{(f)}(x) =0.1$)) and a negative saturated $d$-type GSF (${\cal N}_g^{(d)}(x) =-1$). Shaded area represents a $\pm 20$\% uncertainty on ${\cal N}_g^{(f)}$. Data are from Ref.~\cite{Adare:2013ekj}.}
\label{fig:AN-pi}
\end{center}
\end{figure}

In Fig.~\ref{fig:AN-pi} (left panel) we present the maximized (${\cal N}_g(x) = +1$) gluon Sivers contributions to $A_N$ for the process $p^\uparrow p\to \pi^0\, X$  at $\sqrt s=200$ GeV and mid-rapidity as a function of $p_T$, together with PHENIX data~\cite{Adare:2013ekj}, for the $f$-type (red solid line) and $d$-type (blue dot-dashed line) pieces. For completeness we also show the maximized gluon Sivers term in the GPM (green dashed line).
As mentioned above, the quark Sivers contribution, also within the CGI-GPM scheme and adopting the parametrization as extracted from SIDIS data~\cite{Anselmino:2008sga}, is totally negligible (red dotted line).
From this plot we realize that while the $d$-type contribution, for this process and in this kinematical region, is dynamically suppressed, the $f$-type one can be potentially large. The reason is that for the $d$-type term the hard partonic cross sections for the processes initiated by $gq$ and $g\bar q$ pairs enter with a relative sign (see Eqs.~(\ref{eq:Hgqgq}) and (\ref{eq:Hgqqg})) and at mid-rapidity the quark and anti-quark unpolarized TMD parton distributions are equally important. On top of that, there is no $gg\to gg$ contribution (see Eq.~(\ref{eq:Hgggg})), the dominant channel at moderate values of $p_T$. This is in contrast with the $f$-type term, which indeed could be potentially very large. We also notice that the corresponding effect in the GPM approach is even larger: the reason is that its partonic contributions are exactly those entering the unpolarized cross section, all positive and unsuppressed.

These considerations lead us to the second step of our strategy: the attempt to describe reasonably well the $A_N$ data for $\pi^0$ production at mid-rapidity within the CGI-GPM approach, by adopting at the same time the most conservative (that is less stringent) bounds on the $f$- and $d$-type GSFs. Notice that in the region where they are more precise ($p_T \lesssim 5$ GeV), the data are tiny, of the order of per mille, and positive. It is then clear that the most conservative scenario that could give SSAs comparable to the data implies a cancellation between the two contributions, with a strongly suppressed and positive $f$-type GSF and a saturated, negative $d$-type one (supposed totally unknown).
The corresponding results, for ${\cal N}_g^{(f)}(x)=+0.1$ and ${\cal N}_g^{(d)}(x)=-1$, are shown in the right panel of Fig.~\ref{fig:AN-pi}, together with an estimated overall uncertainty band of about $\pm$20\% on ${\cal N}_g^{(f)}$.
Notice that a smaller $d$-type GSF (in size, that is either positive or negative) would imply an even smaller $f$-type GSF. This issue will be addressed in the following.

\begin{figure}[t]
\begin{center}
\includegraphics[trim = 1.cm 0cm 1cm 0cm, width=8cm]{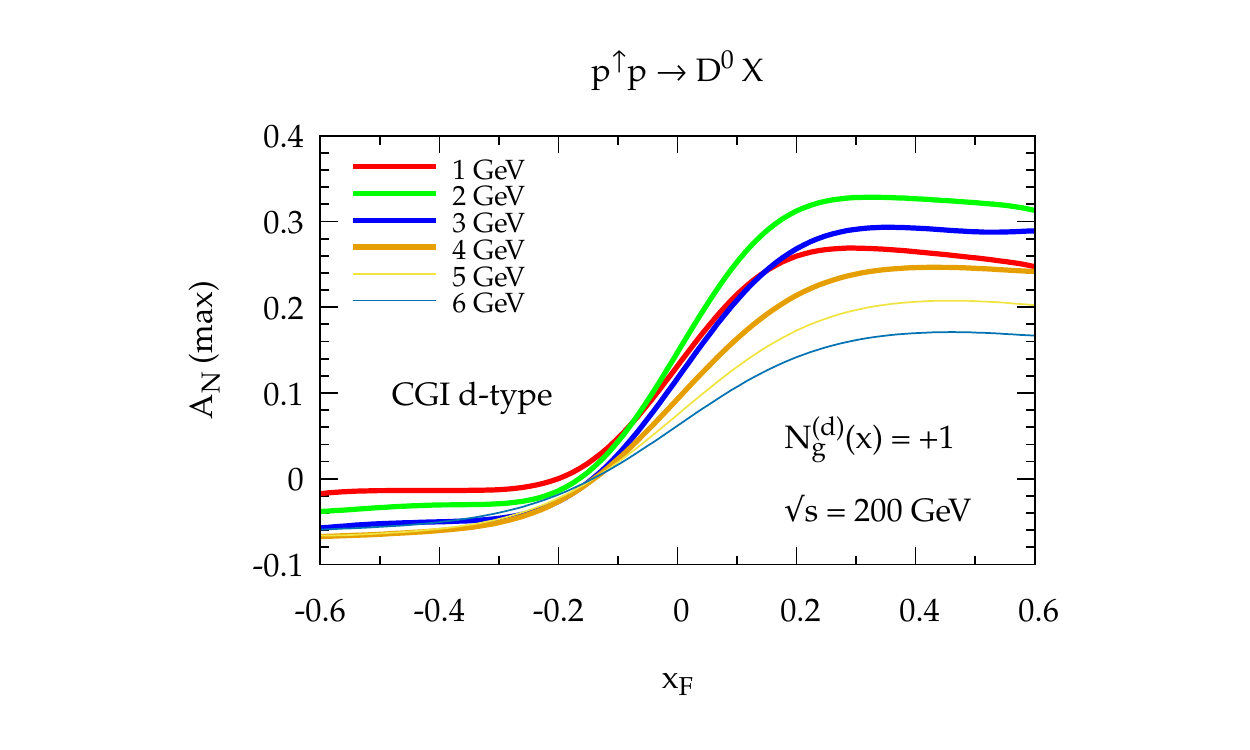}
\includegraphics[trim = 1.cm 0cm 1cm 0cm,width=8cm]{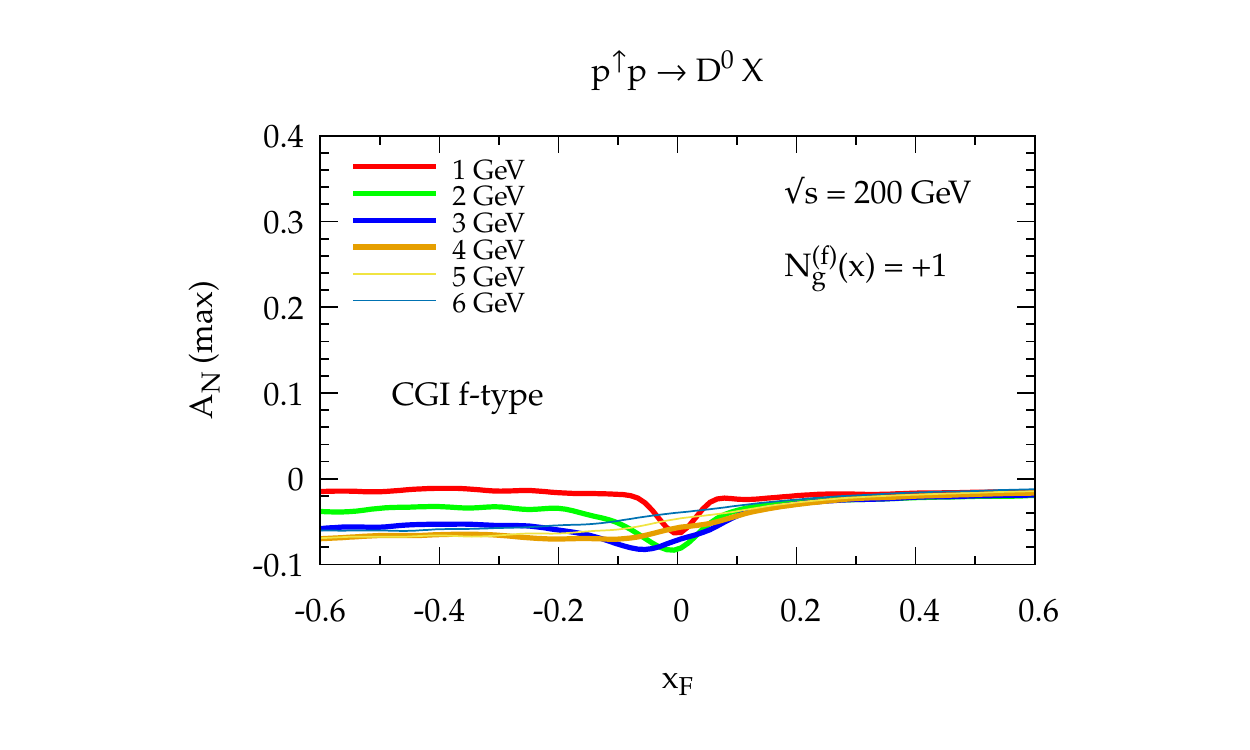}
\caption{Maximized (${\cal N}_g(x) =+1$) $A_N$ for the process $p^\uparrow p\to D^0\, X$ at $\sqrt s=200$ GeV and different $p_T$ values (between 1 and 6 GeV) as a function of $x_F$, within the CGI-GPM approach: $d$-type (left panel) and $f$-type (right panel) contributions.}
\label{fig:AN-D0-sat}
\end{center}
\end{figure}

Let us now consider $A_N$ for $D^0$ production at $\sqrt s= 200$ GeV in the kinematical region relevant to carry out the corresponding analysis for its muon decays, for which data are available~\cite{Aidala:2017pum}. Actually, to be more general, we consider an even larger region both in $x_F=2p_{L}/\sqrt s$ (where $p_L$ is the $D$ meson longitudinal momentum)  and $p_T$. In Fig.~\ref{fig:AN-D0-sat} we show the results for $A_N$ as a function of $x_F$  and for different $p_T$ values, obtained by separately maximizing the $d$- (left panel) and $f$-type (right panel) contributions, as explained above. One can see that in the forward region, while the $d$-type term could be sizeable, the $f$-type one is relatively small. This is in contrast to what was discussed above for the case of $\pi^0$ production. The reason is that, since for $D^0$ production at leading order we consistently consider only the dominant fragmentation of the charm quark into the heavy meson, the cancellations between the $gq$ and $g\bar{q}$ initiated processes, affecting the previous case, are not present anymore. Moreover, the hard partonic parts favor the $d$-type w.r.t.~the $f$-type term: as one can see from Eq.~(41) of Ref.~\cite{DAlesio:2017rzj}, besides some common factors, the hard part for the $f$-type GSF contains a factor $\hat t^2/\hat s^2$, whilst that for the $d$-type GSF contains a term $(\hat t^2-2\hat u^2)/\hat s^2$. Since $|\hat t|$ becomes smaller and smaller as $x_F$ increases, the first piece is relatively suppressed w.r.t.~the second one. On the other hand, in the backward region, where the two hard parts are similar, both contributions are relatively suppressed by the integration over the Sivers azimuthal phase, which for $x_F<0$ is less effective in the hard parts.

If we now use the information extracted from the analysis of $\pi^0$ SSA data, the $f$-type contribution in Fig.~\ref{fig:AN-D0-sat} should be accordingly reduced by a factor of about 0.1 (coming from the corresponding GSF), thus becoming practically negligible.
This implies that for $D^0$ production only the $d$-type GSF could be considered active and therefore constrained by a comparison with the available data. Similar considerations apply also to $\bar D^0$ production. In this case, as discussed in Ref.~\cite{DAlesio:2017rzj}, within the CGI-GPM approach the $f$-type contribution to $A_N$ is the same as the one for $D^0$ production, while the $d$-type gets an opposite sign.

At this point, one has to convert the estimates for $D$-meson production to the corresponding SSAs for its muon decay products\footnote{We thank Jeongsu Bok (PHENIX Collaboration) for providing us with the muon SSA results,
obtained from our $D$-meson estimates.}, for which data are available~\cite{Aidala:2017pum}.
Notice that in our LO approach the SSAs for $D^0$ and $D^+$ production (leading to the $\mu^+$ results) are equal, as are those for $\bar D^0$ and $D^-$ production ($\mu^-$ results).

Since the muon SSA data are still very few and with large error bars, we refrain from performing a fit, and will consider a simple $x$-independent ${\cal N}_g^{(f,d)}(x)\equiv N_g^{(f,d)}$. In the following we discuss different possible scenarios for the $d$- and $f$-type GSFs, taking into account the complementary information coming from $\pi^0$ SSAs. As we will see in a moment, even from this very conservative approach we can extract some important information.

\begin{figure}[t]
\begin{center}
\includegraphics[trim = 1.cm 0cm 1cm 0cm, width=8cm]{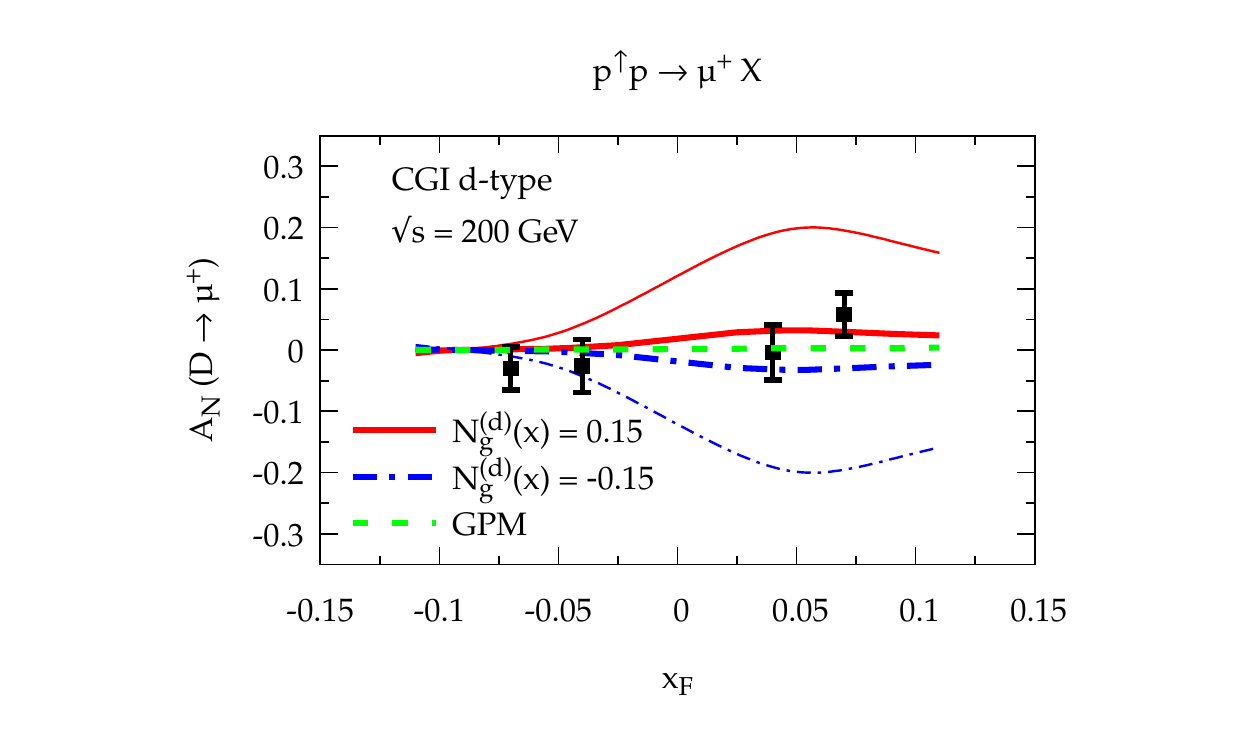}
\includegraphics[trim = 1.cm 0cm 1cm 0cm, width=8cm]{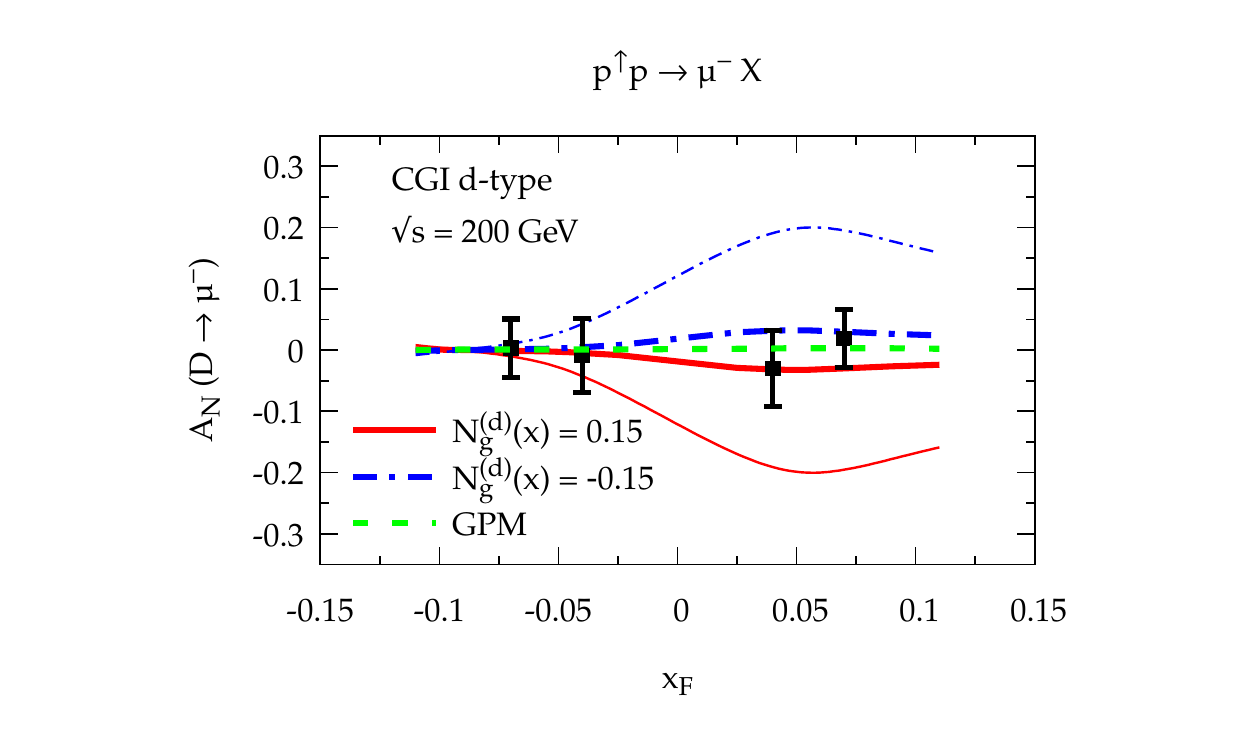}
\caption{$d$-type gluon contributions, within the CGI-GPM approach, to $A_N$ for the process $p^\uparrow p\to \mu^+\, X$ (left panel) and $p^\uparrow p\to \mu^-\, X$ (right panel) from $D$-meson production at $\sqrt s=200$ GeV as a function of $x_F$: maximized effect, ${\cal N}_g^{(d)}(x) =+1$ (thin red solid lines), ${\cal N}_g^{(d)}(x) =-1$ (thin blue dot-dashed lines); reduced (constrained) contribution, ${\cal N}_g^{(d)}(x) =+0.15$ (thick red solid lines), ${\cal N}_g^{(d)}(x) =-0.15$ (thick blue dot-dashed lines). GPM predictions (green dashed lines) are also shown. Data are from Ref.~\cite{Aidala:2017pum}.}
\label{fig:AN-muon}
\end{center}
\end{figure}

\begin{figure}[t]
\begin{center}
\includegraphics[trim = 1.cm 0cm 1cm 0cm, width=8cm]{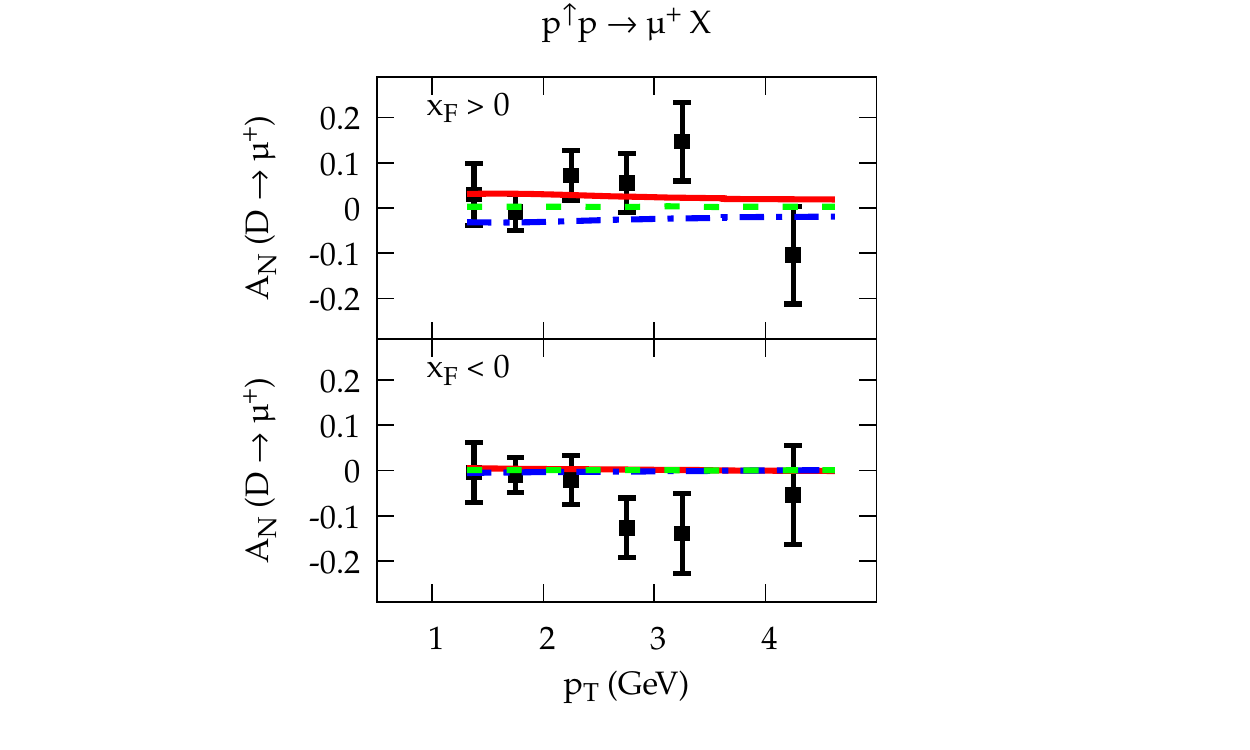}
\includegraphics[trim = 1.cm 0cm 1cm 0cm, width=8cm]{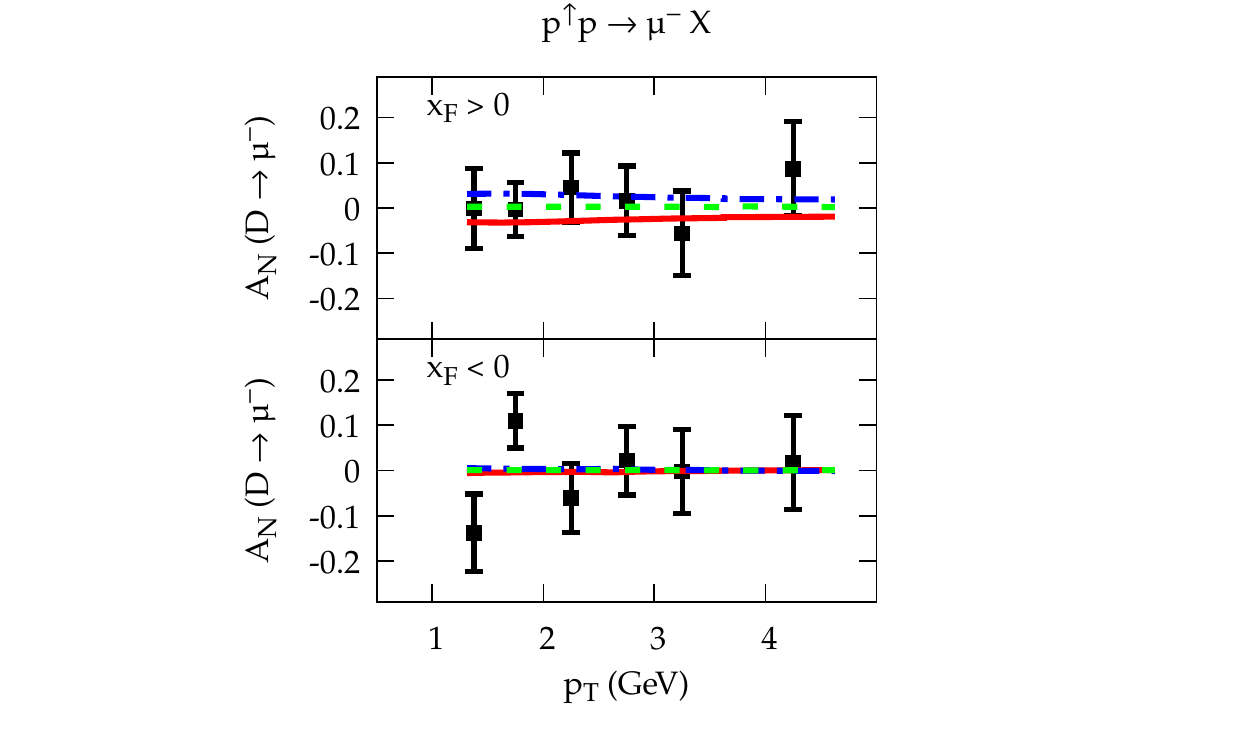}
\caption{$d$-type gluon contributions, within the CGI-GPM approach, to $A_N$ for the process $p^\uparrow p\to \mu^+\, X$ (left panel) and $p^\uparrow p\to \mu^-\, X$ (right panel) from $D$-meson production at $\sqrt s=200$ GeV as a function of $p_T$:  ${\cal N}_g^{(d)}(x) =+0.15$ (thick red solid lines), ${\cal N}_g^{(d)}(x) =-0.15$ (thick blue dot-dashed lines). GPM predictions (green dashed lines) are also shown. Data are from Ref.~\cite{Aidala:2017pum}.}
\label{fig:AN-muonpT}
\end{center}
\end{figure}

As one can see from Fig.~\ref{fig:AN-muon}, for both $\mu^+$ (left panel) and $\mu^-$ (right panel) production, the data are compatible with zero, with only one data point, at the largest Feynman $x$ value, slightly positive for the $\mu^+$ case. It is also clear that the maximized $d$-type GSF contributions (thin red solid line: ${\cal N}_g^{(d)} = +1$, thin blue dot-dashed line: ${\cal N}_g^{(d)} = -1$) largely overestimate the positive $x_F$ experimental data in size. Notice that the value ${\cal N}_g^{(d)} = -1$, together with ${\cal N}_g^{(f)} = + 0.1$, was adopted in order to reasonably reproduce the $\pi^0$ SSA data (see Fig.~\ref{fig:AN-pi}, right panel). On the other hand, to get a fair account of the muon SSA data, one has to take indicatively $|{\cal N}_g^{(d)}| \leq 0.15$, with a mild preference for positive values, because of the positive $\mu^+$ data point. As an example, the results obtained adopting ${\cal N}_g^{(d)} = +0.15(-0.15)$ are shown as thick red solid lines (thick blue dot-dashed lines) in Fig.~\ref{fig:AN-muon} both for $\mu^+$ (left panel) and $\mu^-$ (right panel) production.
Taking into account this new piece of information on the $d$-type GSF, we can reconsider the pion SSA data more accurately.
We find that by varying ${\cal N}_g^{(d)}$ in the range $-0.15\,\div\,+0.15$, while keeping $\rho=2/3$, a very good description of both the $\mu^\pm$ and $\pi^0$ data can be obtained by taking ${\cal N}_g^{(f)}$ in the \emph{corresponding} range $+0.05\,\div\,-0.01$, that is:
\begin{eqnarray}
\label{eq:par_gsf_CGI}
{\cal N}_g^{(d)} = - 0.15\, & \rightarrow & {\cal N}_g^{(f)} = +0.05 \nonumber \\
{\cal N}_g^{(d)} = + 0.15\, & \rightarrow & {\cal N}_g^{(f)} = -0.01\,.
\end{eqnarray}
In other words, a stronger suppression of the $f$-type GSF is required by the combined analysis of muon and pion SSA data.
On the contrary, in the GPM approach the parametrization of the GSF extracted from the $\pi^0$ SSA data, see Eq.~(\ref{eq:par_gsf_GPM}), leads to SSAs for $\mu^\pm$ leptons in very good agreement with available data (Fig.~\ref{fig:AN-muon}, green dashed lines). For completeness, in Fig.~\ref{fig:AN-muonpT} we also show the corresponding SSA estimates as a function of $p_T$ in the positive and negative $x_F$ regions.

It is worth recalling that a similar analysis of SSAs for $D$-meson production, within the twist-three approach, was carried out in Ref.~\cite{Koike:2011mb}. The corresponding predictions for muon production were compared against the data in Ref.~\cite{Aidala:2017pum}, showing a similar fairly good agreement.  

A few comments on the above procedure are in order. The use of a fixed $\rho$ value implies a fixed $k_\perp$ dependence of the GSF, therefore no such information has been extracted within the CGI-GPM approach. On the other hand, the adopted value leaves the size of the GSF practically unconstrained. Then, by tuning the parameter $N_g$ against the data we can control and estimate its size. We have also to remind that there are strong correlations between these parameters, but the amount and the precision of available data, as already stated above, prevent us from performing a true fit.

For all these reasons, in the following we will show only the first $k_\perp$-moment of the GSF, which better represents its size in an almost unbiased form (at least in the $x$ region probed by the data, $10^{-3}\le x \le 0.4$), without speculating on its detailed $k_\perp$ or $x$ dependences. Further studies in this respect will be necessary.

\begin{figure}[t]
\begin{center}
\includegraphics[width=10cm]{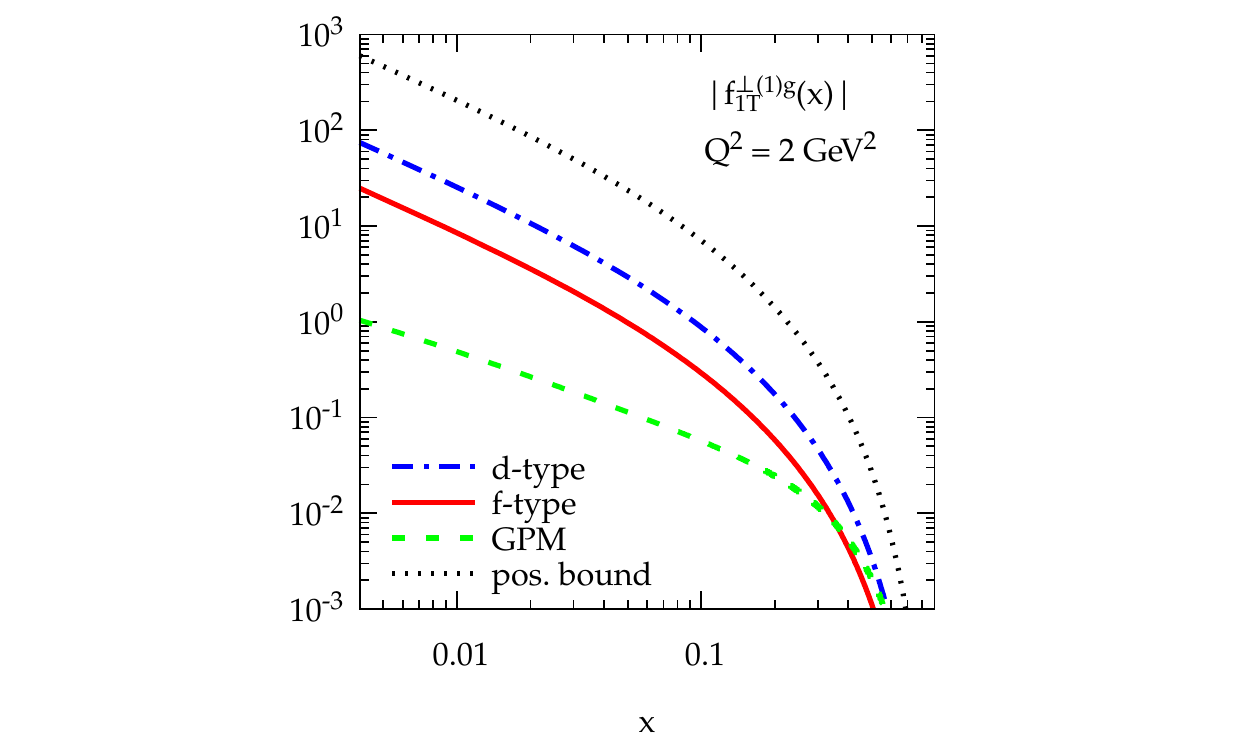}
\caption{Upper values for the first $k_\perp$-moments of the gluon Sivers functions in different approaches and scenarios at $Q^2 = 2$ GeV$^2$: GPM approach (green dashed line), CGI-GPM $d$-type (blue dot-dashed line) and $f$-type ($N_g^{(f)}=0.05$, red solid line). The positivity bound (black dotted line) is also shown.}
\label{fig:1stmom}
\end{center}
\end{figure}

In Fig.~\ref{fig:1stmom} we show the results for the absolute value of the first $k_\perp$-moment of the GSFs as extracted from our  analysis for the GPM (green dashed line) and the CGI-GPM approaches, $d$-type (blue dot-dashed line) and $f$-type ($N_g^{(f)}=0.05$, red solid line), together with the positivity bound (black dotted line). The most stringent bound is the one for the GPM approach, since in this case there are no relative cancellations between the hard partonic parts, being them all positive.
In contrast, the $d$-type GSF within the CGI-GPM scheme is the less bounded (see comments above).

\begin{figure}[t]
\begin{center}
\includegraphics[trim = 1.cm 0cm 1cm 0cm,width=8cm]{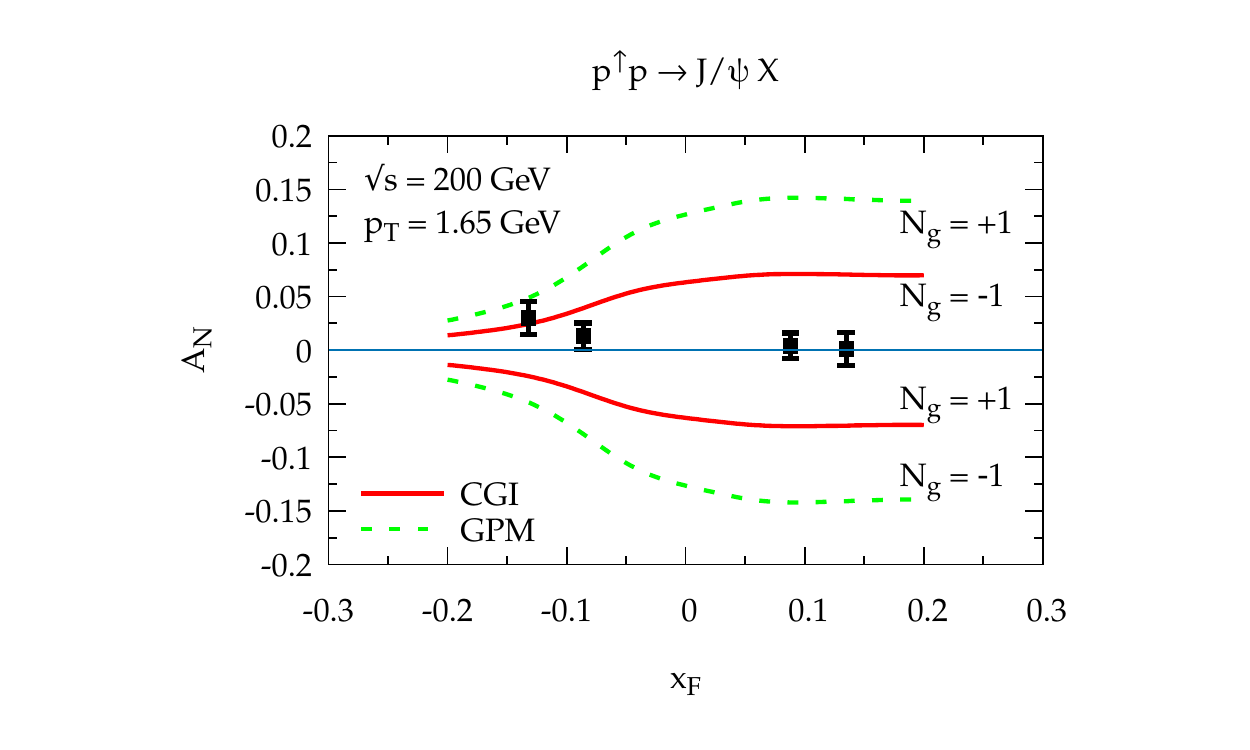}
\includegraphics[trim = 1.cm 0cm 1cm 0cm,width=8cm]{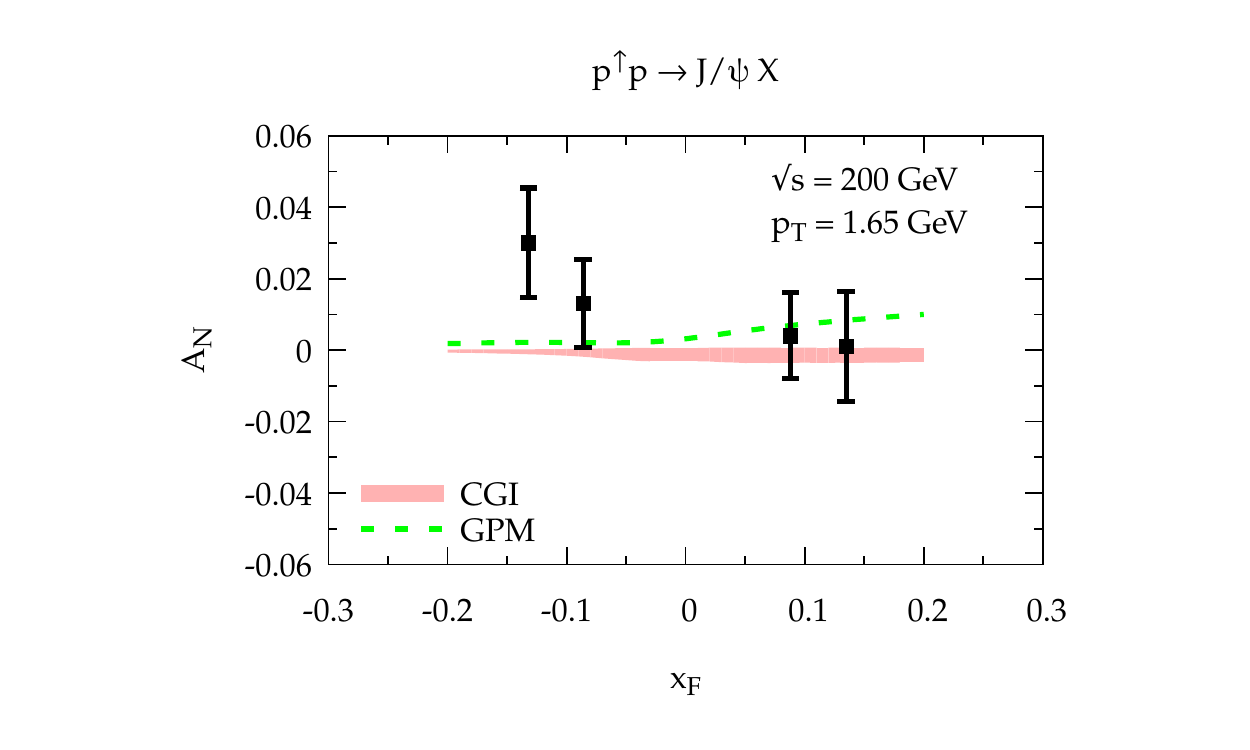}
\caption{$A_N$ for the process $p^\uparrow p\to J/\psi\, X$ at $\sqrt s=200$ GeV and $p_T = 1.65$ GeV as a function of $x_F$.
Left panel: maximized $A_N$ in the GPM (green dashed lines) and the CGI-GPM (red solid lines) approaches adopting $|{\cal N}_g(x)|=~1$. Right panel: estimates within the GPM (green dashed line) and the CGI-GPM approaches, $f$-type (red band), obtained adopting the GSFs from the present analysis (see Eqs.~(\ref{eq:par_gsf_GPM}), (\ref{eq:par_gsf_CGI})). Data are from Ref.~\cite{Aidala:2018gmp}. Notice the different scales in the two panels.}
\label{fig:Jpsi}
\end{center}
\end{figure}

\begin{figure}[t]
\begin{center}
\includegraphics[trim = 1.cm 0cm 1cm 0cm,width=8cm]{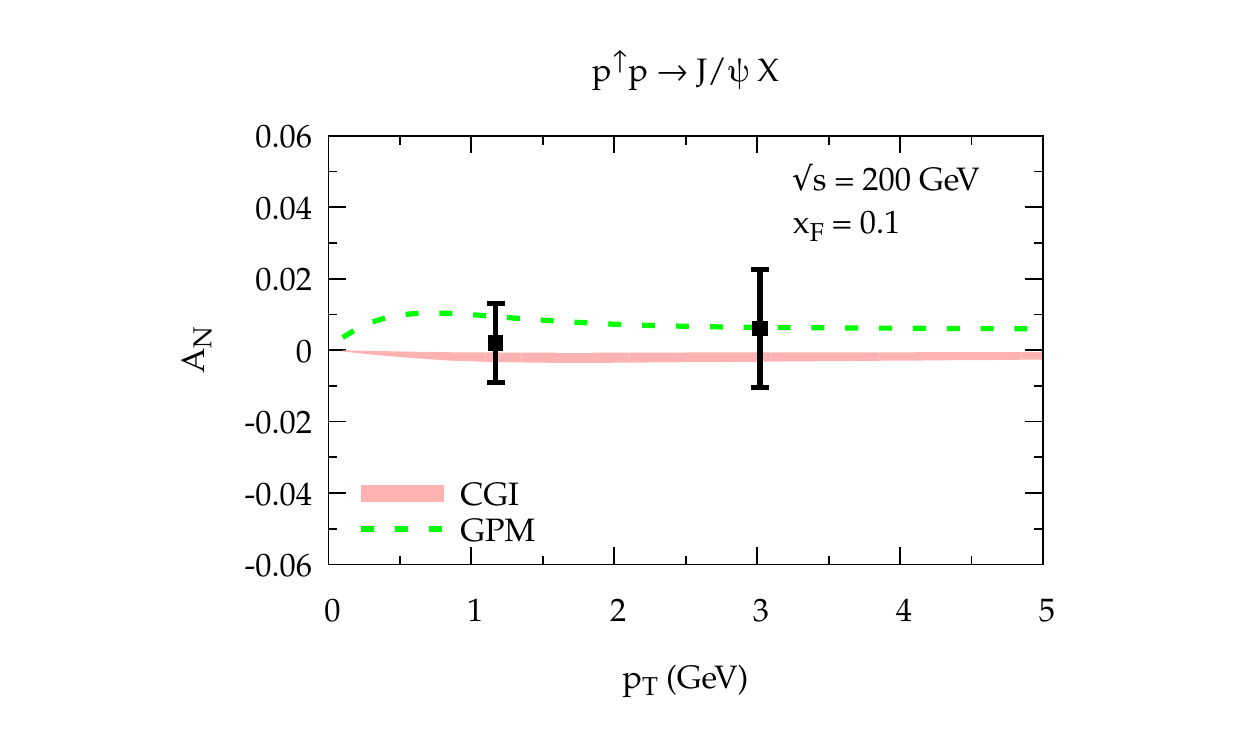}
\includegraphics[trim = 1.cm 0cm 1cm 0cm,width=8cm]{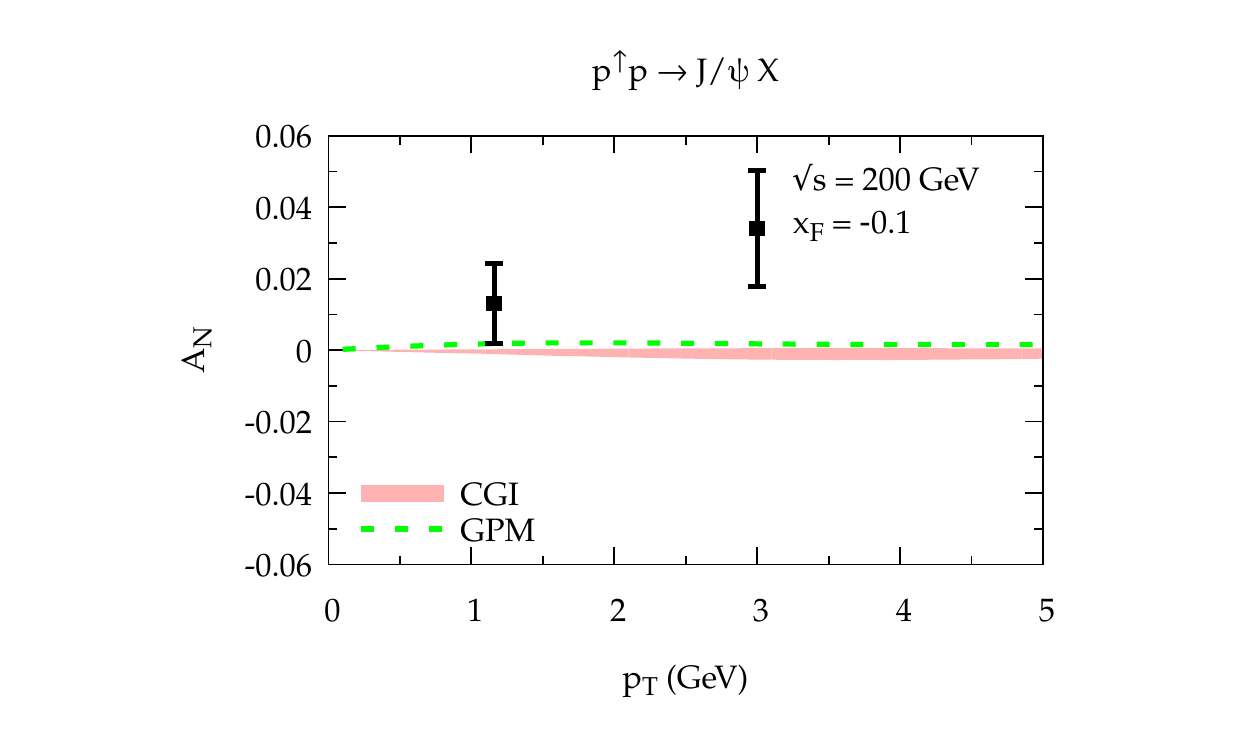}
\caption{Estimates of $A_N$ for the process $p^\uparrow p\to J/\psi\, X$ at $\sqrt s=200$ GeV and $x_F=+0.1$ (left panel) and $x_F=-0.1$ (right panel) as a function of $p_T$ in the GPM (green dashed line) and the CGI-GPM approaches (red band), adopting the GSFs as extracted in the present analysis (see Eqs.~(\ref{eq:par_gsf_GPM}), (\ref{eq:par_gsf_CGI})). Data are from Ref.~\cite{Aidala:2018gmp}.}
\label{fig:Jpsi2}
\end{center}
\end{figure}

\begin{figure}[t]
\begin{center}
\includegraphics[trim = 1.cm 0cm 1cm 0cm,width=8cm]{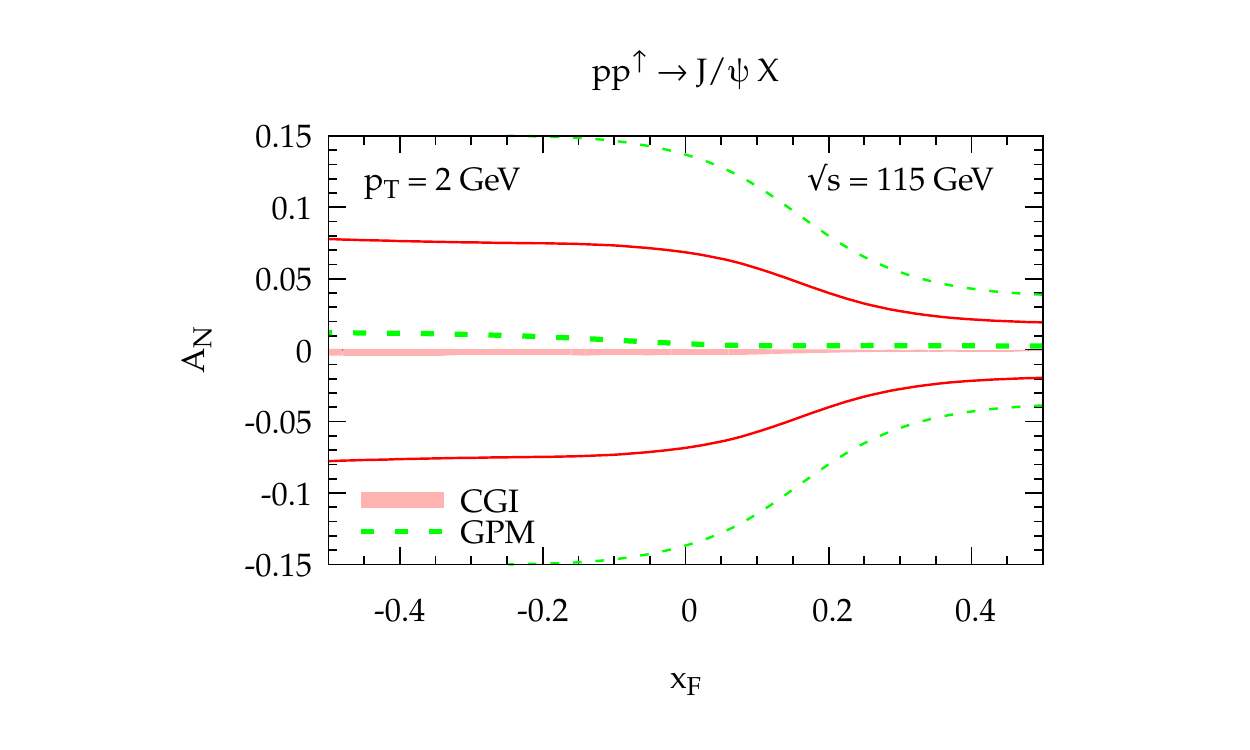}
\includegraphics[trim = 1.cm 0cm 1cm 0cm,width=8cm]{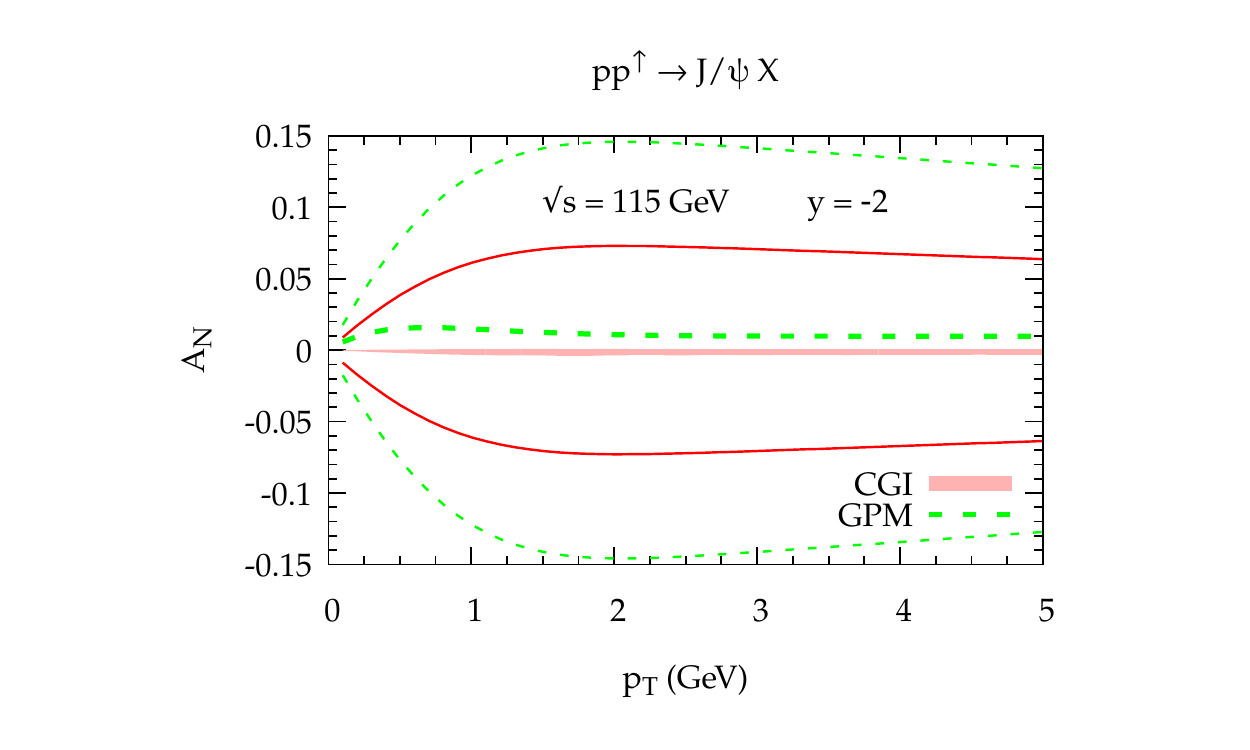}
\caption{$A_N$ for the process $pp^\uparrow \to J/\psi\, X$ at $\sqrt s=115$ GeV and $p_T = 2$ GeV as a function of $x_F$ (left panel) and at rapidity $y=-2$ as a function of $p_T$ (right panel). Notice that here negative rapidities correspond to the forward region for the polarized proton. Predictions are for the GPM (thick green dashed lines) and the CGI-GPM (red band) approaches (see Eqs.~(\ref{eq:par_gsf_GPM}), (\ref{eq:par_gsf_CGI})). The corresponding maximized contributions for the GPM (thin green dashed lines) and the CGI-GPM (red solid lines) schemes are also shown.
}
\label{fig:Jpsi-LHC}
\end{center}
\end{figure}

\begin{figure}[t]
\begin{center}
\includegraphics[trim = 1.cm 0cm 1cm 0cm,width=8cm]{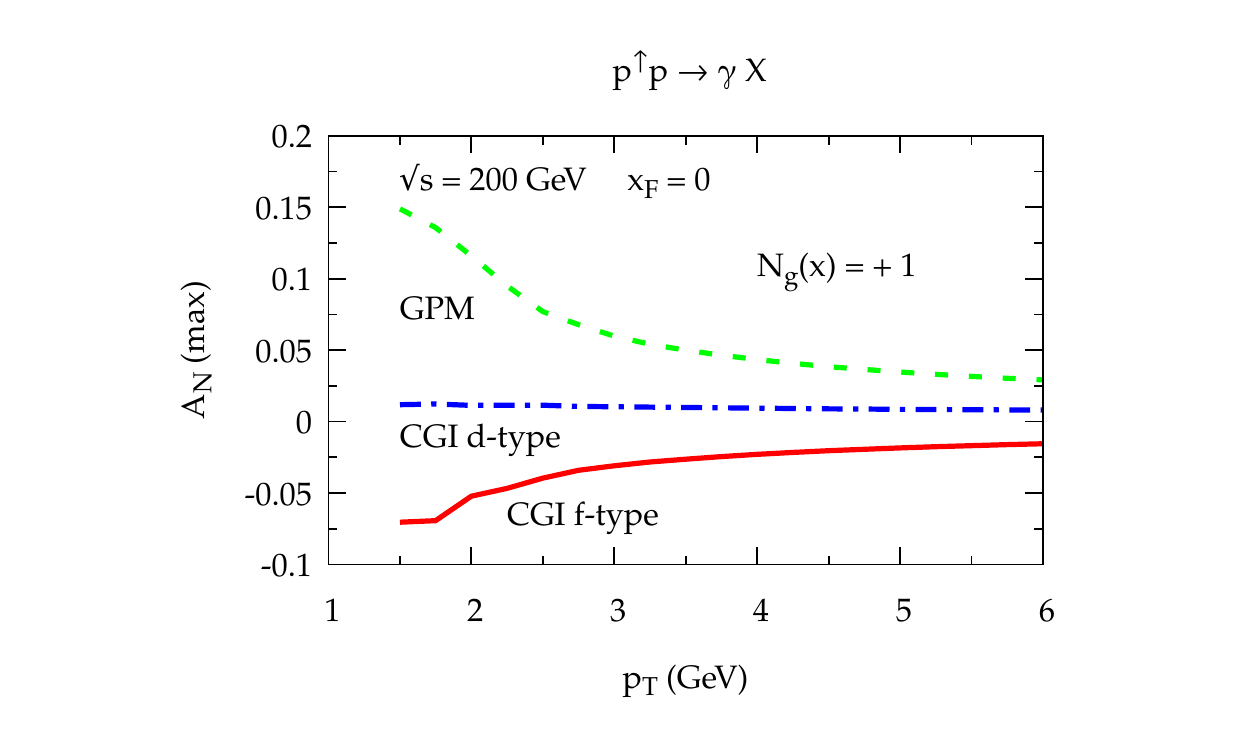}
\includegraphics[trim = 1.cm 0cm 1cm 0cm,width=8cm]{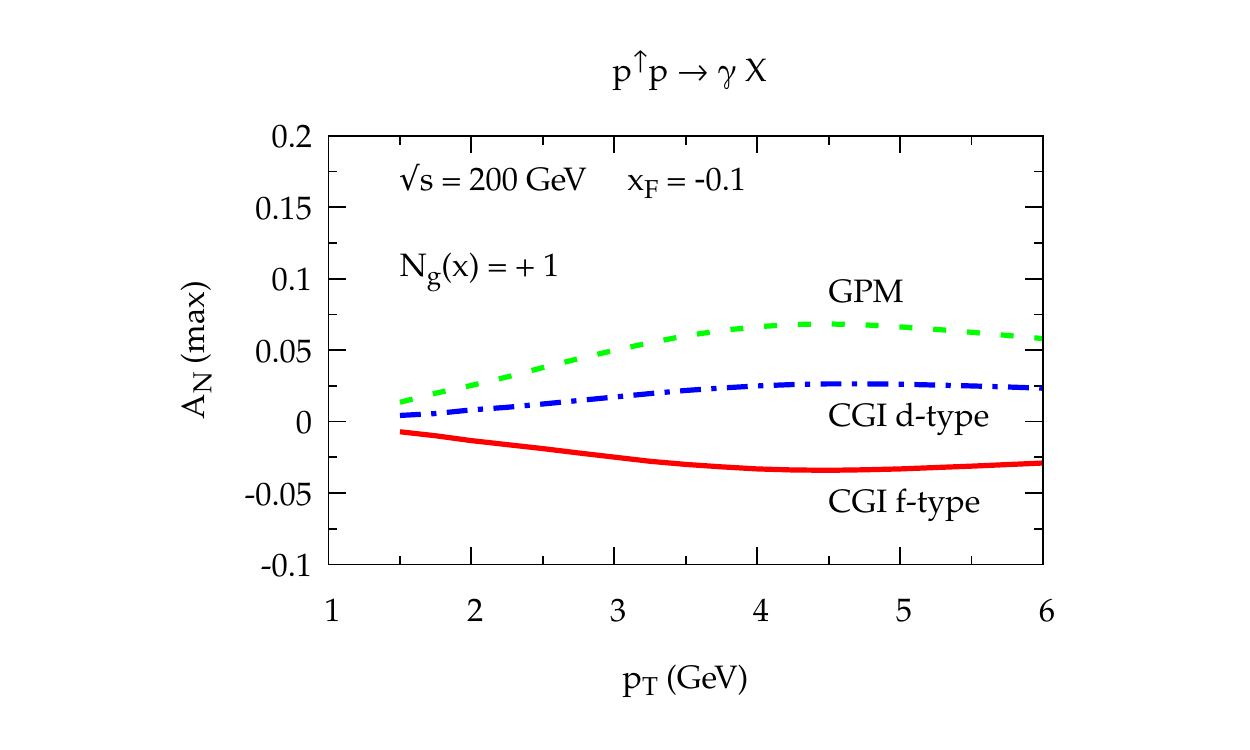}

\includegraphics[trim = 1.cm 0cm 1cm 0cm,width=8cm]{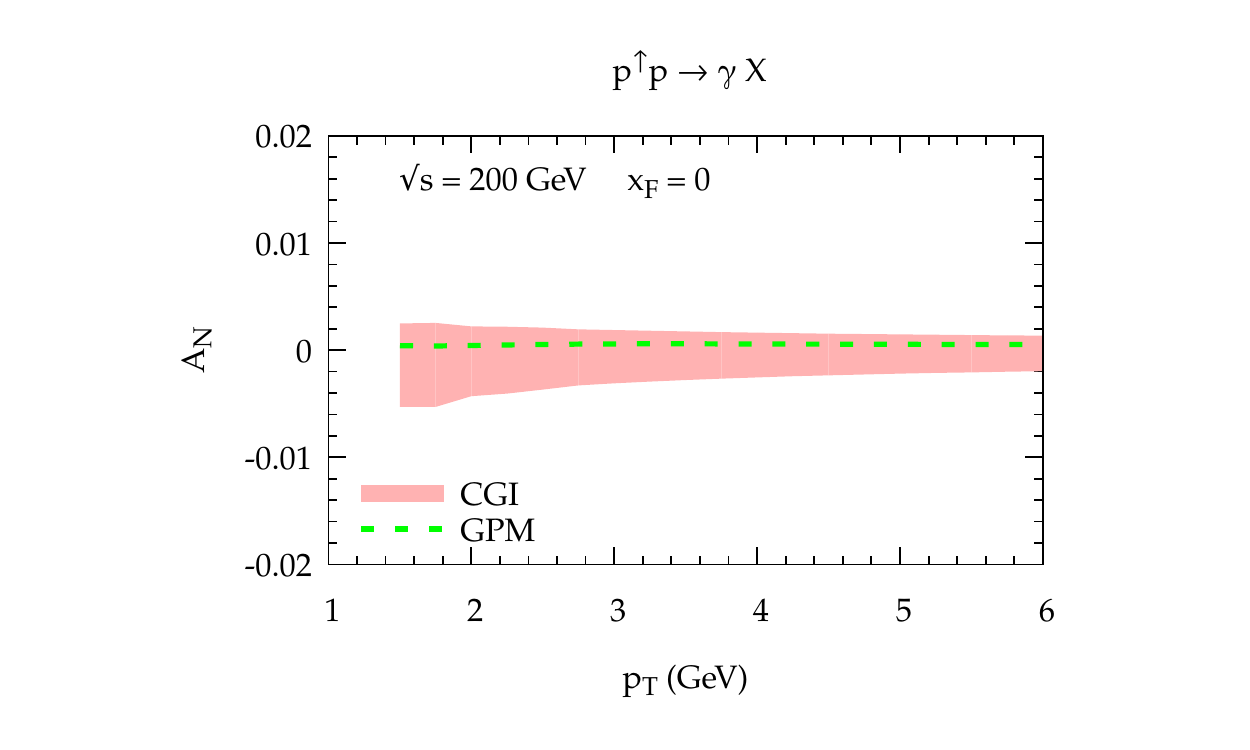}
\includegraphics[trim = 1.cm 0cm 1cm 0cm,width=8cm]{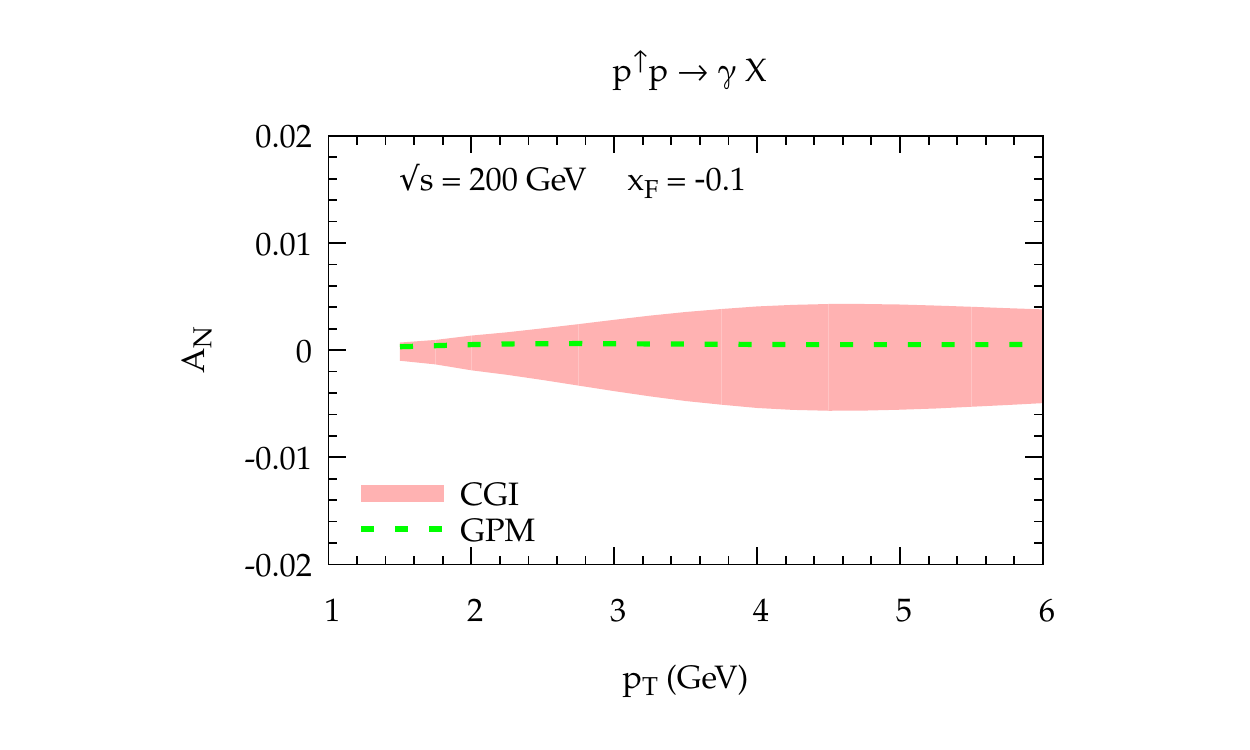}
\caption{Estimates of $A_N$ for the process $p^\uparrow p\to \gamma\, X$ at $\sqrt s=200$ GeV as a function of $p_T$ within the GPM and the CGI-GPM approaches. Upper panels: maximized contributions (${\cal N}_g(x) =+1$) at $x_F=0$ (left) and $x_F=-0.1$ (right); lower panels, estimates based on the present analysis (see Eqs.~(\ref{eq:par_gsf_GPM}), (\ref{eq:par_gsf_CGI})): GPM (green dashed line), CGI-GPM (red band).}
\label{fig:AN-gamma}
\end{center}
\end{figure}

\subsection {Predictions for SSAs in $p^\uparrow p\to J/\psi\, X$ and $p^\uparrow p\to \gamma\, X$}
\label{predict}

As discussed in Ref.~\cite{DAlesio:2017rzj}, $A_N$ for $J/\psi$ production is directly sensitive to the gluon Sivers function. Moreover, within the CGI-GPM approach and the Color Singlet model, only the $f$-type distribution contributes to the Sivers asymmetry. In Figs.~\ref{fig:Jpsi} and \ref{fig:Jpsi2} we show a comparison of our estimates, evaluated adopting $M_T^2= M_{J/\psi}^2+p_T^2$ as factorization scale, with PHENIX data~\cite{Aidala:2018gmp} for $A_N$ in $p^\uparrow p\to J/\psi\, X$. In particular, in Fig.~\ref{fig:Jpsi}, left panel, we show the maximized (${\cal N}_g^{(f)}(x)=\pm 1$) contributions to $A_N$ at fixed $p_T = 1.65$ GeV as a function of $x_F$, both in the GPM (green dashed lines) and the CGI-GPM (red solid lines) approaches. Notice that, also in this case, the integration over the Sivers azimuthal phase strongly suppresses the SSA in the backward-rapidity region. In the right panel of Fig.~\ref{fig:Jpsi} we present our corresponding predictions based on the present analysis: GPM (green dashed line), CGI-GPM (red band: $-0.01\le {\cal N}_g^{(f)} \le 0.05$).
In Fig.~\ref{fig:Jpsi2} we show the corresponding estimates as a function of $p_T$ at $x_F=0.1$ (left panel) and $x_F=-0.1$ (right panel).

With the exception of the experimental point in the most backward rapidity region, data are compatible with zero and our estimates describe them fairly well. Notice that, in principle, by using a suitable $x$-dependent factor, ${\cal N}_g^{(f)}(x)$ (namely something like $N_g\,(1-x)^\beta$, with $N_g \simeq -1$ and a large $\beta$), also the data points at $x_F<0$ could be accounted for. On the other hand, this would prevent a description of pion SSAs at small $p_T$, which require a strong suppression of the $f$-type GSF, in particular in the small-$x$ region (see Fig.~\ref{fig:AN-pi}, left panel). If $J/\psi$ measurements would be confirmed even in future higher statistics samples, this would definitely represent a tension with the pion SSAs, at least within a TMD approach. In this respect, more data, on a wider kinematical range and with better statistics, would be very helpful.

It is worth considering the corresponding analysis for $A_N$ in $J/\psi$ production for the kinematics reachable at LHC in the fixed target mode with a transversely polarized target (see the AFTER~\cite{Brodsky:2012vg,Hadjidakis:2018ifr} and LHCb~\cite{LHCbspin:2017xx,LHCbspin:2018xx} proposals at CERN). In such a configuration one could probe even larger light-cone momentum fractions in the polarized proton, accessing the gluon TMDs in a very interesting and complementary region.

In Fig.~\ref{fig:Jpsi-LHC} we present our estimates for $A_N$ for $p p^\uparrow \to J/\psi\, X$ at $\sqrt s=115$ GeV, at fixed $p_T=2$~GeV, as a function of $x_F$ (left panel) and at fixed rapidity $y=-2$, as a function of $p_T$ (right panel). Notice that in such a configuration the backward rapidity region refers to the forward region for the polarized proton target. In particular, we show our predictions within the GPM (thick green dashed lines) and the CGI-GPM (red bands) approaches, together with the corresponding upper/lower positivity bounds (thin lines). From these results we see that any further experimental information would be extremely useful.

Another interesting observable, where the gluon Sivers function could be directly accessed, is the SSA in $p^\uparrow p \to \gamma\, X$, for which we have given the complete expressions in the CGI-GPM scheme in the previous Section. We present here some estimates, both in the GPM and CGI-GPM approaches, saturating their contributions as well as adopting the results of the phenomenological analysis presented above. As for the case of SSAs in $\pi^0$ production, the most interesting regions are those at mid- and slightly backward-rapidity and not so large values of $|x_F|$. The reason is that, at large negative values of $x_F$, the integration over the Sivers azimuthal phase washes out the effect.
This would partially spoil the analysis proposed in Ref.~\cite{{Schmidt:2005gv}}, where the authors discussed this process as a clear tool to access the GSF, also in this kinematical region.

In Fig.~\ref{fig:AN-gamma} (upper panels) we show the maximized contributions to $A_N$ at $x_F=0$ (left) and $x_F=-0.1$ (right). As one can see, the $d$-type term at $x_F=0$ is dynamically suppressed, as for the $\pi^0$ production case: the reason is indeed the same, that is the partial cancellation between the hard $gq\to\gamma q$ and $g\bar{q}\to\gamma\bar{q}$ processes, see Eq.~(\ref{eq:Hgqgammaq2}). Indeed, this suppression is less pronounced at $x_F=-0.1$, where the unpolarized quark and anti-quark TMDs inside the unpolarized proton are probed at larger $x$ values and therefore are not equally important. 
Moreover, in the small $p_T$ range (up to 3 GeV) the maximized estimates at $x_F=-0.1$ are more suppressed w.r.t.~those at $x_F=0$, due, once again, to the integration over the Sivers azimuthal phase. In the lower panels we show our estimates adopting the results discussed in the previous subsection. In all cases the values are very small and compatible with zero. Despite of this, a measure of $A_N$ for direct photon production would be extremely important to test the consistency of the whole approach.

\section{Conclusions}
\label{concl}

In this paper we have performed a study of the gluon Sivers function through a combined analysis of data on transverse single-spin asymmetries for the processes $p^\uparrow p \to \pi^0\, X$~\cite{Adare:2013ekj} and $p^\uparrow p \to D\, X \to \mu\, X$~\cite{Aidala:2017pum}, measured by the PHENIX Collaboration at RHIC. The theoretical framework adopted is the so-called transverse momentum dependent generalized parton model (GPM), in which intrinsic parton motion and spin effects are considered. In addition, we have used the color gauge invariant version of this model (CGI-GPM), which takes into account also, in the one-gluon exchange approximation, the initial and final state interactions of the active parton with the remnants of the polarized proton, leading to a process dependent Sivers function.

From a theoretical point of view, we have extended the calculation of the expressions for the single-spin asymmetries in  $p^\uparrow p \to \pi\, X$ and  $p^\uparrow p \to \gamma\, X$, within the CGI-GPM approach, to the gluon sector. In this way, we completed the study of Ref.~\cite{Gamberg:2010tj}, in which only the corresponding quark-induced subprocesses were studied.
As a byproduct, we have also shown that the one-gluon approximation employed here is sufficient to recover the exact gluonic pole strengths in any partonic process calculated at LO in perturbative QCD~\cite{Bomhof:2006ra} (see the Appendices).

The analogous formulae for the single spin asymmetries in $p^\uparrow p \to D\, X$ and $p^\uparrow p \to J/\psi\, X$ were derived in Ref.~\cite{DAlesio:2017rzj}. It turns out that for these processes the gluon Sivers function can be re-expressed as a linear combination of two independent, universal (and so far unknown) contributions, namely the $f$-type and $d$-type Sivers distributions.

On the phenomenological side, using available knowledge of the quark and antiquark Sivers functions from SIDIS measurements, we have shown how the PHENIX data on inclusive pion and $D$-meson production allow us to partially disentangle and considerably constrain the size of these two gluon Sivers functions, which should be much smaller than their positivity bounds. This can be considered the first significant attempt towards a quantitative extraction of these process dependent gluon Sivers functions.
On the other hand, since the number and the precision of the available data is not very high, our findings have still to be considered as preliminary.
Furthermore, we have compared the extractions of the gluon Sivers function in the two approaches, with (CGI-GPM scheme) and without (GPM scheme) initial/final state interactions. The results are encouraging, even if it is not yet possible to clearly discriminate between the GPM and the CGI-GPM frameworks.

Our results have been used to predict the single-spin asymmetry for the processes $p^\uparrow p\to J/\psi\,X$, which only depends on the $f$-type Sivers function. Comparison with existing PHENIX data~\cite{Aidala:2018gmp}, compatible with zero at forward rapidities, shows a good agreement. Predictions for the same processes have been presented in a kinematic region accessible at LHC with a fixed polarized target, and for the process $p^\uparrow p \to \gamma\, X$ at RHIC kinematics as well, for which data are not yet available. These will certainly help in shedding light on the still poorly known gluon Sivers function and towards our understanding of the three-dimensional structure of the nucleons.\\

Note added: at the very last stage of this work we have become aware of a similar study on SSAs in $p^\uparrow p\to \gamma\, X$ within the CGI-GPM approach~\cite{Godbole:2018mmh}. While the theoretical findings are in perfect agreement, see Eqs.~\eqref{eq:Hgqgammaq1}-\eqref{eq:Hgqgammaq2}, the phenomenological analysis presents some differences, which deserve further attention. A possible explanation could be the different way of handling the role of the azimuthal phases (to be integrated over in the final observable) in the hard partonic pieces.

\acknowledgments
We would like to thank Jeongsu Bok (PHENIX Collaboration) for providing us with the conversion of our estimates for $D$-meson SSAs into the corresponding ones for muon production. We also acknowledge useful discussion with Chen Xu concerning the analysis of the SSA data in $J/\psi$ production. The work of P.T.\ is supported by the European Research Council (ERC) under the European Union's Horizon 2020 research and innovation program (grant agreement No.~647981, 3DSPIN). U.D.~acknowledges partial support by Fondazione Sardegna under the project ``Quarkonium at LHC energies", CUP F71I17000160002 (University of Cagliari).

\appendix

\section{Color factors for the gluon Sivers effect in $p^\uparrow p \to \pi\, X$}
\label{sec:cf-pion}
In this Appendix we present the color factors needed for the evaluation of the partonic hard functions $H_{ab\to cd}^{\text{Inc}}$ in the expression of the single spin asymmetry for the process  $p^\uparrow p \to \pi\, X$ in the  CGI-GPM framework. We list the explicit results for the subprocesses $gq\to gq$ (Table~\ref{tab:gq2gq}), $gg \to q \bar{q}$ (Table~\ref{tab:gg2qqb}) and $gg\to gg$ (Table~\ref{tab:gg2gg}).
In all the tables, $C_U$ denotes the usual unpolarized color factor for the specific diagram $D$, while $C_I$, $C_{F_c}$, $C_{F_d}$ are the color factors obtained when an extra gluon is attached in $D$ to parton $b$ ($C_I$), parton $c$ ($C_{F_c}$) or parton $d$  ($C_{F_d}$). Furthermore, for each diagram we need to distinguish between the two possible ways in which color is neutralized, leading to the two possible gluon Sivers functions, $f$-type and $d$-type. For each process, the sum of all diagrams, taken with the new color factors $C_I^{(f/d)}$ and $C^{(f/d)}_{F_c}$, gives $H^{(f/d)}_I$ and $H_{F_c}^{(f/d)}$, respectively, and
\begin{equation}
H^{\text{Inc} \, (f/d)} = H_I^{(f/d)} + H_{F_c}^{(f/d)}.
\end{equation}
Notice that the $C_{F_d}$ factors sum up to zero and do not play any role in the single-inclusive hadron production.

Alternatively, $H^{\text{Inc} \, (f/d)}$ can be obtained directly by summing the diagrams with the color factors
\begin{equation}
C^{\text{Inc}\,(f/d)} \equiv C_I^{(f/d)} + C_{F_c}^{(f/d)}~.
\end{equation}
Finally, we have checked that, for each diagram $D$, the gluonic pole strengths defined by
\begin{equation}
C_G^{(f/d)} = \frac{C_I^{(f/d)} + C_{F_c}^{(f/d)} + C_{F_d}^{(f/d)}}{C_U}\,,
\end{equation}
are in full agreement with the ones given in Ref.~\cite{Bomhof:2006ra} for less inclusive processes like
$p^\uparrow p \to \pi\,\pi\,X$, for which the FSIs of parton $d$ need to be taken into account as well.

\begin{table}[t]
\begin{center}
\includegraphics[trim= 40 500 20 80,clip,width=17cm]{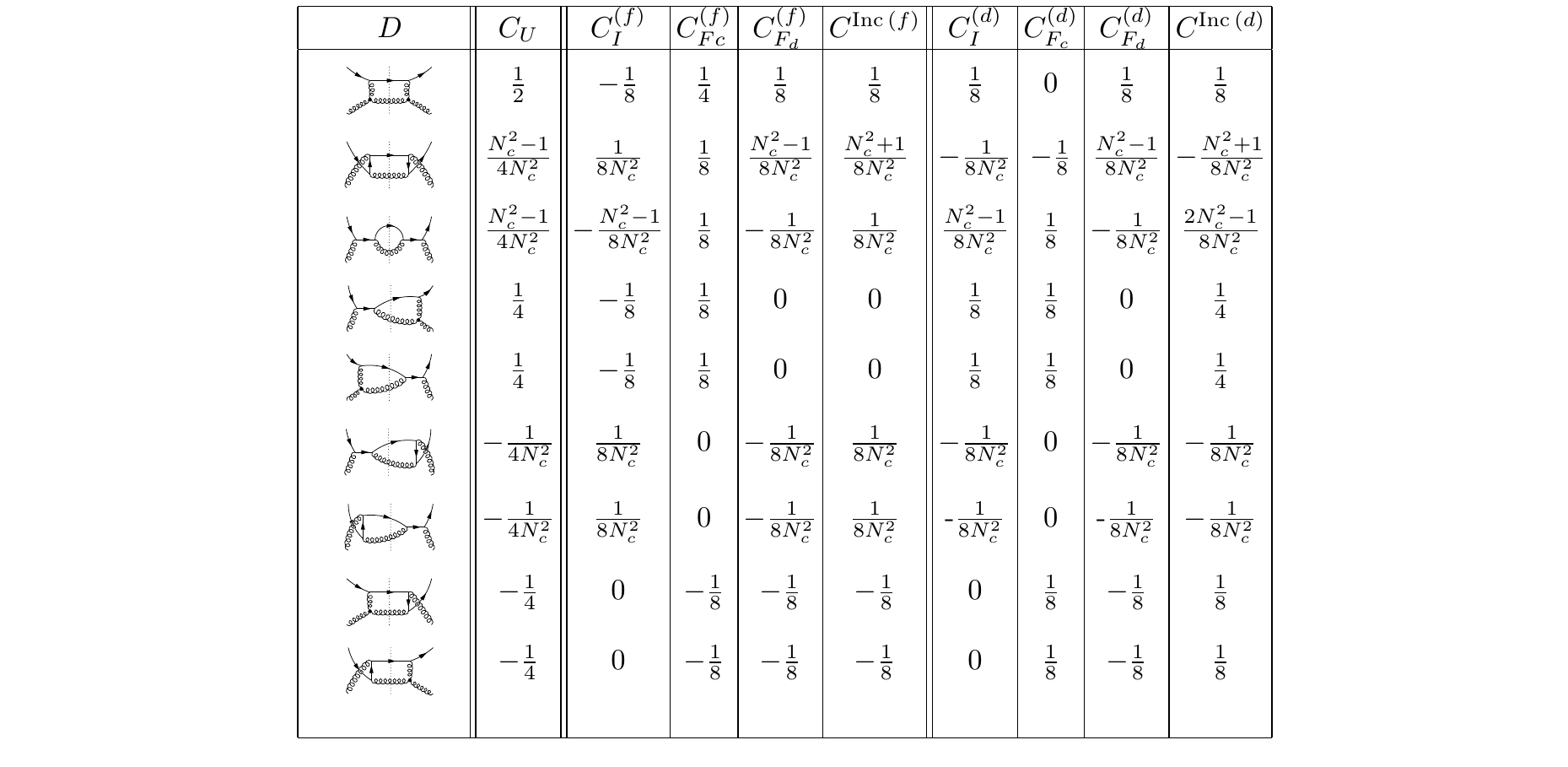}
\end{center}
\caption{Color factors for the LO diagrams contributing to the process $gq\to g q$. $C_U$ denotes the unpolarized color factor for the diagram $D$, while $C_I$, $C_{F_c}$ and $C_{F_d}$, respectively for the $f$- and $d$-type, are the color factors obtained when an extra gluon is attached in $D$ to parton $b$ ($C_I$), parton $c$ ($C_{F_c}$) or parton $d$  ($C_{F_d}$). Furthermore, $C^{\rm Inc} = C_I+C_{F_c}$.}
\label{tab:gq2gq}
\end{table}

\begin{table}[t]
\begin{center}
\includegraphics[trim= 40 480 0 80,clip,width=17cm]{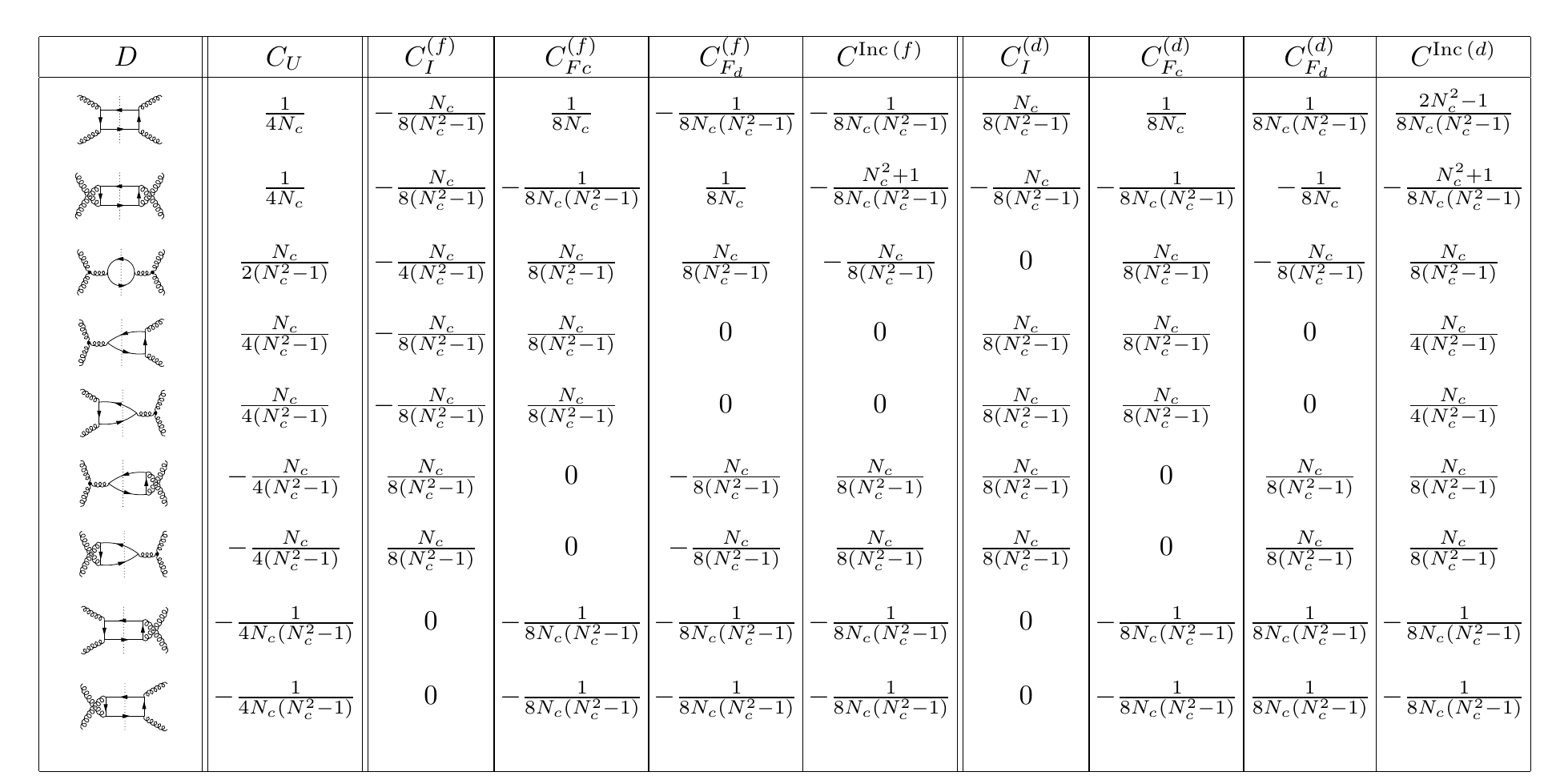}
\end{center}
\caption{Color factors for the LO diagrams contributing to the process $gg\to q\bar q$. Notation is the same as in Tab.~\ref{tab:gq2gq}.}
\label{tab:gg2qqb}
\end{table}

\begin{table}[t]
\begin{center}
\includegraphics[trim= 40 500 20 80,clip,width=17cm]{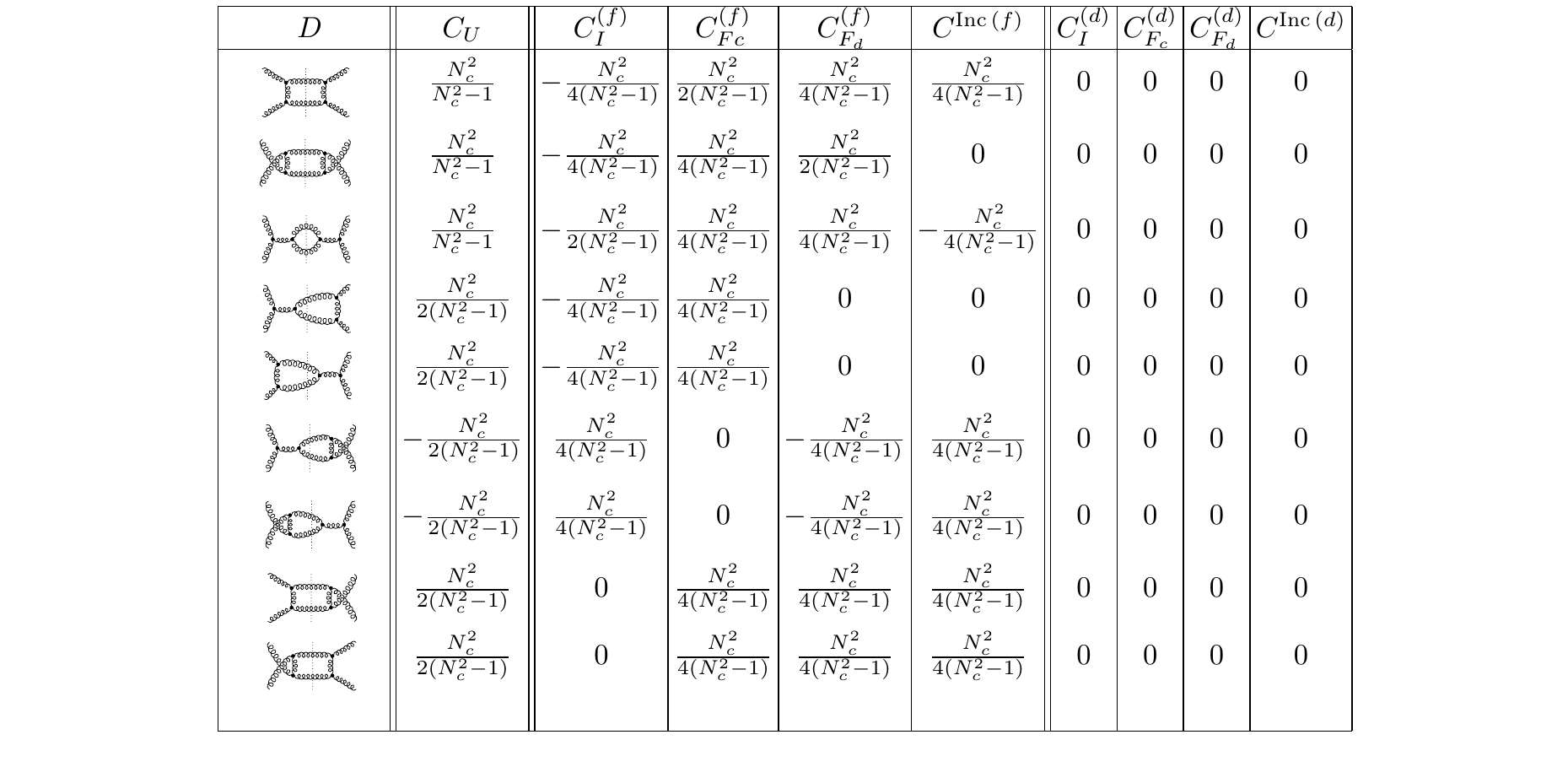}
\end{center}
\caption{Color factors for the LO diagrams contributing to the process $gg\to gg$. Notation is the same as in Tab.~\ref{tab:gq2gq}. In this case all $C^{(d)}$ color factors are zero for symmetry reasons.}
\label{tab:gg2gg}
\end{table}

\section{Color factors for the gluon Sivers effect in $p^\uparrow p \to \gamma\, X$}
 \label{sec:cf-gamma}

The hard functions needed for the calculation of the Sivers effect in $p^\uparrow p \to \gamma\, X$, evaluated in the framework of the CGI-GPM, are given by
\begin{equation}
H^{\text{Inc}\, (f/d)}_{ab\to \gamma d }= \frac{C^{\text{Inc}\, (f/d)}}{C_U}H^{U}_{ab\to \gamma d } \equiv  \frac{C_I^{(f/d)}}{C_U}H^{U}_{ab\to \gamma d } \,,
\end{equation}
where $ab\to \gamma d$ is a generic partonic subprocess contributing to $p^\uparrow p \to \gamma\, X$. Our results for the color factors relevant for the gluon induced subprocesses $gq\to \gamma q$ and $g\bar{q}\to \gamma \bar{q}$ are summarized in Table~\ref{tab}.  Due to their simple color structures, all diagrams $D$ have the same color factors.  As before, $C_U$ is the unpolarized one, while $C_I $ $(C_{Fd})$ is the color factor obtained when an extra gluon is attached in $D$ to parton $b$ (parton $d$).
Since the photon does not interact with the remnant of the polarized nucleon,  $C_{F_c}^{(f/d)} =0$.

Finally, we point out that our {\it gluonic pole strengths}, defined as
\begin{equation}
C_G^{(f/d)} \equiv \frac{C_I^{(f/d)}  + C^{(f/d)}_{F_d}}{C_U}\,,
\end{equation}
are in full agreement with the ones given in Table B.4 of Ref.~\cite{Bomhof:2007zz} for $g q\to \gamma q$, namely
\begin{equation}
C_G^{(f)} = 0\,\qquad C_G^{(d)} = 1~.
\end{equation}

\begin{table}[t]
\begin{center}
\includegraphics[trim= 50 670 20 80,clip,width=17cm]{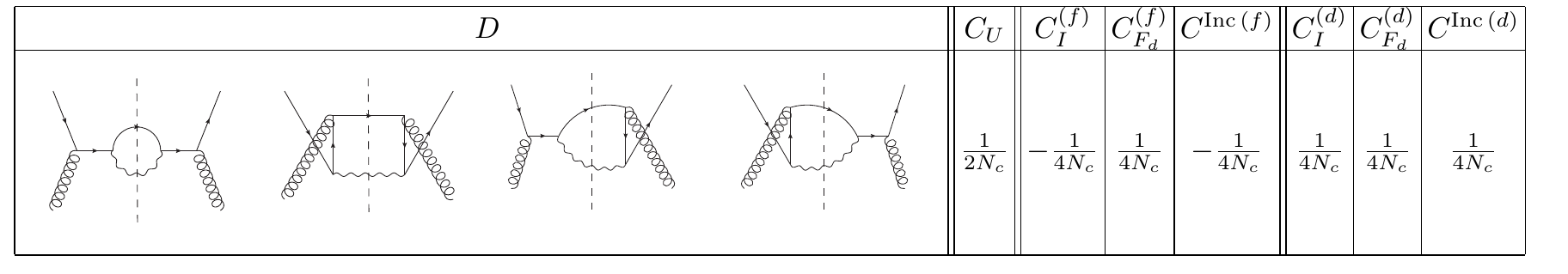}
\end{center}
 \caption{Color factors for the process $g q \to \gamma q$. For the process $g \bar{q} \to \gamma \bar{q}$, the $f$-type color factors are the same, while the $d$-ones have an overall minus sign. Notation is the same as in Tab.~\ref{tab:gq2gq}.}
\label{tab}
\end{table}

Notice that the results in Ref.~\cite{Bomhof:2007zz} have been derived adopting a different method, {\it i.e.}~by looking at the full gauge link structure and taking the derivative of the gauge link.

For completeness, the hard functions for the quark induced subprocesses, calculated in Ref.~\cite{Gamberg:2010tj}, are presented  below:
 \begin{align}
 H^{\text{Inc}}_{qg\to \gamma q}  & = -H^{\text{Inc}}_{\bar{q}g\to \gamma \bar{q}}  = -\frac{N_c^2-1}{N_c^2}\left  ( -\frac{\hat t}{\hat s} - \frac{\hat s}{\hat t}   \right )\,, \nonumber \\
 H^{\text{Inc}}_{q \bar q \to \gamma g}  & = - H^{\text{Inc}}_{\bar q  q \to \gamma g} = \frac{1}{N_c^2}\,\left ( \frac{\hat t}{\hat u} + \frac{\hat u}{\hat t} \right )\,.
 \end{align}


\begin{thebibliography}{46}
\expandafter\ifx\csname natexlab\endcsname\relax\def\natexlab#1{#1}\fi
\expandafter\ifx\csname bibnamefont\endcsname\relax
  \def\bibnamefont#1{#1}\fi
\expandafter\ifx\csname bibfnamefont\endcsname\relax
  \def\bibfnamefont#1{#1}\fi
\expandafter\ifx\csname citenamefont\endcsname\relax
  \def\citenamefont#1{#1}\fi
\expandafter\ifx\csname url\endcsname\relax
  \def\url#1{\texttt{#1}}\fi
\expandafter\ifx\csname urlprefix\endcsname\relax\def\urlprefix{URL }\fi
\providecommand{\bibinfo}[2]{#2}
\providecommand{\eprint}[2][]{\url{#2}}

\bibitem[{\citenamefont{Sivers}(1990)}]{Sivers:1989cc}
\bibinfo{author}{\bibfnamefont{D.~W.} \bibnamefont{Sivers}},
  \bibinfo{journal}{Phys. Rev.} \textbf{\bibinfo{volume}{D41}},
  \bibinfo{pages}{83} (\bibinfo{year}{1990}).

\bibitem[{\citenamefont{Sivers}(1991)}]{Sivers:1990fh}
\bibinfo{author}{\bibfnamefont{D.~W.} \bibnamefont{Sivers}},
  \bibinfo{journal}{Phys. Rev.} \textbf{\bibinfo{volume}{D43}},
  \bibinfo{pages}{261} (\bibinfo{year}{1991}).

\bibitem[{\citenamefont{Collins}(2002)}]{Collins:2002kn}
\bibinfo{author}{\bibfnamefont{J.~C.} \bibnamefont{Collins}},
  \bibinfo{journal}{Phys. Lett.} \textbf{\bibinfo{volume}{B536}},
  \bibinfo{pages}{43} (\bibinfo{year}{2002}), \eprint{hep-ph/0204004}.

\bibitem[{\citenamefont{Brodsky et~al.}(2002)\citenamefont{Brodsky, Hwang, and
  Schmidt}}]{Brodsky:2002rv}
\bibinfo{author}{\bibfnamefont{S.~J.} \bibnamefont{Brodsky}},
  \bibinfo{author}{\bibfnamefont{D.~S.} \bibnamefont{Hwang}}, \bibnamefont{and}
  \bibinfo{author}{\bibfnamefont{I.}~\bibnamefont{Schmidt}},
  \bibinfo{journal}{Nucl. Phys.} \textbf{\bibinfo{volume}{B642}},
  \bibinfo{pages}{344} (\bibinfo{year}{2002}), \eprint{hep-ph/0206259}.

\bibitem[{\citenamefont{Airapetian et~al.}(2005)}]{Airapetian:2004tw}
\bibinfo{author}{\bibfnamefont{A.}~\bibnamefont{Airapetian}}
  \bibnamefont{et~al.} (\bibinfo{collaboration}{HERMES Collaboration}),
  \bibinfo{journal}{Phys. Rev. Lett.} \textbf{\bibinfo{volume}{94}},
  \bibinfo{pages}{012002} (\bibinfo{year}{2005}), \eprint{hep-ex/0408013}.

\bibitem[{\citenamefont{Adolph et~al.}(2012)}]{Adolph:2012sp}
\bibinfo{author}{\bibfnamefont{C.}~\bibnamefont{Adolph}} \bibnamefont{et~al.}
  (\bibinfo{collaboration}{COMPASS Collaboration})
  \bibinfo{journal}{Phys. Lett.} \textbf{\bibinfo{volume}{B717}},
  \bibinfo{pages}{383} (\bibinfo{year}{2012}),
  \eprint{1205.5122}.

\bibitem[{\citenamefont{Burkardt}(2004)}]{Burkardt:2004ur}
\bibinfo{author}{\bibfnamefont{M.}~\bibnamefont{Burkardt}},
  \bibinfo{journal}{Phys. Rev.} \textbf{\bibinfo{volume}{D69}},
  \bibinfo{pages}{091501} (\bibinfo{year}{2004}), \eprint{hep-ph/0402014}.

\bibitem[{\citenamefont{Anselmino
  et~al.}(2005{\natexlab{a}})\citenamefont{Anselmino, Boglione, D'Alesio, Kotzinian, Murgia, and Prokudin}}]{Anselmino:2005ea}
\bibinfo{author}{\bibfnamefont{M.}~\bibnamefont{Anselmino}},
\bibinfo{author}{\bibfnamefont{M.}~\bibnamefont{Boglione}},
\bibinfo{author}{\bibfnamefont{U.}~\bibnamefont{D'Alesio}},
\bibinfo{author}{\bibfnamefont{A.}~\bibnamefont{Kotzinian}},
\bibinfo{author}{\bibfnamefont{F.}~\bibnamefont{Murgia}},
\bibnamefont{and}
  \bibinfo{author}{\bibfnamefont{A.}~\bibnamefont{Prokudin}},
  \bibinfo{journal}{Phys. Rev.}
  \textbf{\bibinfo{volume}{D72}}, \bibinfo{pages}{094007}
  (\bibinfo{year}{2005}{\natexlab{a}}), \eprint{hep-ph/0507181}.

\bibitem[{\citenamefont{Anselmino et~al.}(2009{\natexlab{a}})\citenamefont{Anselmino, Boglione, D'Alesio, Kotzinian, Melis, Murgia, and Prokudin}}]{Anselmino:2008sga}
\bibinfo{author}{\bibfnamefont{M.}~\bibnamefont{Anselmino}},
\bibinfo{author}{\bibfnamefont{M.}~\bibnamefont{Boglione}},
\bibinfo{author}{\bibfnamefont{U.}~\bibnamefont{D'Alesio}},
\bibinfo{author}{\bibfnamefont{A.}~\bibnamefont{Kotzinian}},
\bibinfo{author}{\bibfnamefont{S.}~\bibnamefont{Melis}},
\bibinfo{author}{\bibfnamefont{F.}~\bibnamefont{Murgia}},
\bibinfo{author}{\bibfnamefont{A.}~\bibnamefont{Prokudin}},
\bibnamefont{and}
  \bibinfo{author}{\bibfnamefont{C.}~\bibnamefont{T\"urk}},
 \bibinfo{journal}{Eur. Phys. J.}
  \textbf{\bibinfo{volume}{A39}}, \bibinfo{pages}{89} (\bibinfo{year}{2009}),
  \eprint{0805.2677}.

\bibitem[{\citenamefont{Efremov et~al.}(2005)\citenamefont{Efremov, Goeke,
  Menzel, Metz, and Schweitzer}}]{Efremov:2004tp}
\bibinfo{author}{\bibfnamefont{A.~V.} \bibnamefont{Efremov}},
  \bibinfo{author}{\bibfnamefont{K.}~\bibnamefont{Goeke}},
  \bibinfo{author}{\bibfnamefont{S.}~\bibnamefont{Menzel}},
  \bibinfo{author}{\bibfnamefont{A.}~\bibnamefont{Metz}}, \bibnamefont{and}
  \bibinfo{author}{\bibfnamefont{P.}~\bibnamefont{Schweitzer}},
  \bibinfo{journal}{Phys. Lett.} \textbf{\bibinfo{volume}{B612}},
  \bibinfo{pages}{233} (\bibinfo{year}{2005}), \eprint{hep-ph/0412353}.

\bibitem[{\citenamefont{Boer et~al.}(2015)\citenamefont{Boer, Lorc\'e, Pisano,
  and Zhou}}]{Boer:2015vso}
\bibinfo{author}{\bibfnamefont{D.}~\bibnamefont{Boer}},
  \bibinfo{author}{\bibfnamefont{C.}~\bibnamefont{Lorc\'e}},
  \bibinfo{author}{\bibfnamefont{C.}~\bibnamefont{Pisano}}, \bibnamefont{and}
  \bibinfo{author}{\bibfnamefont{J.}~\bibnamefont{Zhou}},
  \bibinfo{journal}{Adv. High Energy Phys.} \textbf{\bibinfo{volume}{2015}},
  \bibinfo{pages}{371396} (\bibinfo{year}{2015}), \eprint{1504.04332}.

\bibitem[{\citenamefont{Adare et~al.}(2014)}]{Adare:2013ekj}
\bibinfo{author}{\bibfnamefont{A.}~\bibnamefont{Adare}} \bibnamefont{et~al.}
  (\bibinfo{collaboration}{PHENIX Collaboration}), \bibinfo{journal}{Phys. Rev.}
  \textbf{\bibinfo{volume}{D90}}, \bibinfo{pages}{012006}
  (\bibinfo{year}{2014}), \eprint{1312.1995}.

\bibitem[{\citenamefont{D'Alesio et~al.}(2015)\citenamefont{D'Alesio, Murgia,
  and Pisano}}]{DAlesio:2015fwo}
\bibinfo{author}{\bibfnamefont{U.}~\bibnamefont{D'Alesio}},
  \bibinfo{author}{\bibfnamefont{F.}~\bibnamefont{Murgia}}, \bibnamefont{and}
  \bibinfo{author}{\bibfnamefont{C.}~\bibnamefont{Pisano}},
  \bibinfo{journal}{JHEP} \textbf{\bibinfo{volume}{09}}, \bibinfo{pages}{119}
  (\bibinfo{year}{2015}), \eprint{1506.03078}.

\bibitem[{\citenamefont{Anselmino et~al.}(1995)\citenamefont{Anselmino,
  Boglione, and Murgia}}]{Anselmino:1994tv}
\bibinfo{author}{\bibfnamefont{M.}~\bibnamefont{Anselmino}},
  \bibinfo{author}{\bibfnamefont{M.}~\bibnamefont{Boglione}}, \bibnamefont{and}
  \bibinfo{author}{\bibfnamefont{F.}~\bibnamefont{Murgia}},
  \bibinfo{journal}{Phys. Lett.} \textbf{\bibinfo{volume}{B362}},
  \bibinfo{pages}{164} (\bibinfo{year}{1995}), \eprint{hep-ph/9503290}.

\bibitem[{\citenamefont{D'Alesio and Murgia}(2004)}]{D'Alesio:2004up}
\bibinfo{author}{\bibfnamefont{U.}~\bibnamefont{D'Alesio}} \bibnamefont{and}
  \bibinfo{author}{\bibfnamefont{F.}~\bibnamefont{Murgia}},
  \bibinfo{journal}{Phys. Rev.} \textbf{\bibinfo{volume}{D70}},
  \bibinfo{pages}{074009} (\bibinfo{year}{2004}), \eprint{hep-ph/0408092}.

\bibitem[{\citenamefont{Anselmino
  et~al.}(2006{\natexlab{a}})\citenamefont{Anselmino, Boglione, D'Alesio, Leader, Melis, and Murgia}}]{Anselmino:2005sh}
\bibinfo{author}{\bibfnamefont{M.}~\bibnamefont{Anselmino}},
\bibinfo{author}{\bibfnamefont{M.}~\bibnamefont{Boglione}},
\bibinfo{author}{\bibfnamefont{U.}~\bibnamefont{D'Alesio}},
\bibinfo{author}{\bibfnamefont{E.}~\bibnamefont{Leader}},
\bibinfo{author}{\bibfnamefont{S.}~\bibnamefont{Melis}}, \bibnamefont{and}
  \bibinfo{author}{\bibfnamefont{F.}~\bibnamefont{Murgia}},
\bibinfo{journal}{Phys. Rev.}
  \textbf{\bibinfo{volume}{D73}}, \bibinfo{pages}{014020}
  (\bibinfo{year}{2006}{\natexlab{a}}), \eprint{hep-ph/0509035}.

\bibitem[{\citenamefont{D'Alesio and Murgia}(2008)}]{DAlesio:2007bjf}
\bibinfo{author}{\bibfnamefont{U.}~\bibnamefont{D'Alesio}} \bibnamefont{and}
  \bibinfo{author}{\bibfnamefont{F.}~\bibnamefont{Murgia}},
  \bibinfo{journal}{Prog. Part. Nucl. Phys.} \textbf{\bibinfo{volume}{61}},
  \bibinfo{pages}{394} (\bibinfo{year}{2008}), \eprint{0712.4328}.

\bibitem[{\citenamefont{Anselmino et~al.}(2013)\citenamefont{Anselmino,
  Boglione, D'Alesio, Melis, Murgia, and Prokudin}}]{Anselmino:2013rya}
\bibinfo{author}{\bibfnamefont{M.}~\bibnamefont{Anselmino}},
  \bibinfo{author}{\bibfnamefont{M.}~\bibnamefont{Boglione}},
  \bibinfo{author}{\bibfnamefont{U.}~\bibnamefont{D'Alesio}},
  \bibinfo{author}{\bibfnamefont{S.}~\bibnamefont{Melis}},
  \bibinfo{author}{\bibfnamefont{F.}~\bibnamefont{Murgia}}, \bibnamefont{and}
  \bibinfo{author}{\bibfnamefont{A.}~\bibnamefont{Prokudin}},
  \bibinfo{journal}{Phys. Rev.} \textbf{\bibinfo{volume}{D88}},
  \bibinfo{pages}{054023} (\bibinfo{year}{2013}), \eprint{1304.7691}.

\bibitem[{\citenamefont{Aschenauer et~al.}(2016)\citenamefont{Aschenauer,
  D'Alesio, and Murgia}}]{Aschenauer:2015ndk}
\bibinfo{author}{\bibfnamefont{E.~C.} \bibnamefont{Aschenauer}},
  \bibinfo{author}{\bibfnamefont{U.}~\bibnamefont{D'Alesio}}, \bibnamefont{and}
  \bibinfo{author}{\bibfnamefont{F.}~\bibnamefont{Murgia}},
  \bibinfo{journal}{Eur. Phys. J.} \textbf{\bibinfo{volume}{A52}},
  \bibinfo{pages}{156} (\bibinfo{year}{2016}), \eprint{1512.05379}.

\bibitem[{\citenamefont{D'Alesio
  et~al.}(2017{\natexlab{a}})\citenamefont{D'Alesio, Flore, and
  Murgia}}]{DAlesio:2017nrd}
\bibinfo{author}{\bibfnamefont{U.}~\bibnamefont{D'Alesio}},
  \bibinfo{author}{\bibfnamefont{C.}~\bibnamefont{Flore}}, \bibnamefont{and}
  \bibinfo{author}{\bibfnamefont{F.}~\bibnamefont{Murgia}},
  \bibinfo{journal}{Phys. Rev.} \textbf{\bibinfo{volume}{D95}},
  \bibinfo{pages}{094002} (\bibinfo{year}{2017}{\natexlab{a}}),
  \eprint{1701.01148}.

\bibitem[{\citenamefont{Gamberg and Kang}(2011)}]{Gamberg:2010tj}
\bibinfo{author}{\bibfnamefont{L.}~\bibnamefont{Gamberg}} \bibnamefont{and}
  \bibinfo{author}{\bibfnamefont{Z.-B.} \bibnamefont{Kang}},
  \bibinfo{journal}{Phys. Lett.} \textbf{\bibinfo{volume}{B696}},
  \bibinfo{pages}{109} (\bibinfo{year}{2011}), \eprint{1009.1936}.

\bibitem[{\citenamefont{D'Alesio
  et~al.}(2011{\natexlab{a}})\citenamefont{D'Alesio, Gamberg, Kang, Murgia, and
  Pisano}}]{DAlesio:2011kkm}
\bibinfo{author}{\bibfnamefont{U.}~\bibnamefont{D'Alesio}},
  \bibinfo{author}{\bibfnamefont{L.}~\bibnamefont{Gamberg}},
  \bibinfo{author}{\bibfnamefont{Z.-B.} \bibnamefont{Kang}},
  \bibinfo{author}{\bibfnamefont{F.}~\bibnamefont{Murgia}}, \bibnamefont{and}
  \bibinfo{author}{\bibfnamefont{C.}~\bibnamefont{Pisano}},
  \bibinfo{journal}{Phys. Lett.} \textbf{\bibinfo{volume}{B704}},
  \bibinfo{pages}{637} (\bibinfo{year}{2011}{\natexlab{a}}),
  \eprint{1108.0827}.

\bibitem[{\citenamefont{D'Alesio et~al.}(2014)\citenamefont{D'Alesio, Murgia,
  and Pisano}}]{DAlesio:2013cfy}
\bibinfo{author}{\bibfnamefont{U.}~\bibnamefont{D'Alesio}},
  \bibinfo{author}{\bibfnamefont{F.}~\bibnamefont{Murgia}}, \bibnamefont{and}
  \bibinfo{author}{\bibfnamefont{C.}~\bibnamefont{Pisano}},
  \bibinfo{journal}{Phys. Part. Nucl.} \textbf{\bibinfo{volume}{45}},
  \bibinfo{pages}{676} (\bibinfo{year}{2014}), \eprint{1307.4880}.

\bibitem[{\citenamefont{D'Alesio
  et~al.}(2017{\natexlab{b}})\citenamefont{D'Alesio, Murgia, Pisano, and
  Taels}}]{DAlesio:2017rzj}
\bibinfo{author}{\bibfnamefont{U.}~\bibnamefont{D'Alesio}},
  \bibinfo{author}{\bibfnamefont{F.}~\bibnamefont{Murgia}},
  \bibinfo{author}{\bibfnamefont{C.}~\bibnamefont{Pisano}}, \bibnamefont{and}
  \bibinfo{author}{\bibfnamefont{P.}~\bibnamefont{Taels}},
  \bibinfo{journal}{Phys. Rev.} \textbf{\bibinfo{volume}{D96}},
  \bibinfo{pages}{036011} (\bibinfo{year}{2017}{\natexlab{b}}),
  \eprint{1705.04169}.

\bibitem[{\citenamefont{Bomhof and Mulders}(2007)}]{Bomhof:2006ra}
\bibinfo{author}{\bibfnamefont{C.~J.} \bibnamefont{Bomhof}} \bibnamefont{and}
  \bibinfo{author}{\bibfnamefont{P.~J.} \bibnamefont{Mulders}},
  \bibinfo{journal}{JHEP} \textbf{\bibinfo{volume}{02}}, \bibinfo{pages}{029}
  (\bibinfo{year}{2007}), \eprint{hep-ph/0609206}.

\bibitem[{\citenamefont{Buffing et~al.}(2013)\citenamefont{Buffing, Mukherjee,
  and Mulders}}]{Buffing:2013kca}
\bibinfo{author}{\bibfnamefont{M.~G.~A.} \bibnamefont{Buffing}},
  \bibinfo{author}{\bibfnamefont{A.}~\bibnamefont{Mukherjee}},
  \bibnamefont{and} \bibinfo{author}{\bibfnamefont{P.~J.}
  \bibnamefont{Mulders}}, \bibinfo{journal}{Phys. Rev.}
  \textbf{\bibinfo{volume}{D88}}, \bibinfo{pages}{054027}
  (\bibinfo{year}{2013}), \eprint{1306.5897}.

\bibitem[{\citenamefont{Anselmino et~al.}(2004)\citenamefont{Anselmino,
  Boglione, D'Alesio, Leader, and Murgia}}]{Anselmino:2004nk}
\bibinfo{author}{\bibfnamefont{M.}~\bibnamefont{Anselmino}},
  \bibinfo{author}{\bibfnamefont{M.}~\bibnamefont{Boglione}},
  \bibinfo{author}{\bibfnamefont{U.}~\bibnamefont{D'Alesio}},
  \bibinfo{author}{\bibfnamefont{E.}~\bibnamefont{Leader}}, \bibnamefont{and}
  \bibinfo{author}{\bibfnamefont{F.}~\bibnamefont{Murgia}},
  \bibinfo{journal}{Phys. Rev.} \textbf{\bibinfo{volume}{D70}},
  \bibinfo{pages}{074025} (\bibinfo{year}{2004}), \eprint{hep-ph/0407100}.

\bibitem[{\citenamefont{Godbole et~al.}(2016)\citenamefont{Godbole, Kaushik,
  and Misra}}]{Godbole:2016tvq}
\bibinfo{author}{\bibfnamefont{R.~M.} \bibnamefont{Godbole}},
  \bibinfo{author}{\bibfnamefont{A.}~\bibnamefont{Kaushik}}, \bibnamefont{and}
  \bibinfo{author}{\bibfnamefont{A.}~\bibnamefont{Misra}},
  \bibinfo{journal}{Phys. Rev.} \textbf{\bibinfo{volume}{D94}},
  \bibinfo{pages}{114022} (\bibinfo{year}{2016}), \eprint{1606.01818}.

\bibitem[{\citenamefont{Godbole et~al.}(2017)\citenamefont{Godbole, Kaushik,
  Misra, Rawoot, and Sonawane}}]{Godbole:2017syo}
\bibinfo{author}{\bibfnamefont{R.~M.} \bibnamefont{Godbole}},
  \bibinfo{author}{\bibfnamefont{A.}~\bibnamefont{Kaushik}},
  \bibinfo{author}{\bibfnamefont{A.}~\bibnamefont{Misra}},
  \bibinfo{author}{\bibfnamefont{V.}~\bibnamefont{Rawoot}}, \bibnamefont{and}
  \bibinfo{author}{\bibfnamefont{B.}~\bibnamefont{Sonawane}},
  \bibinfo{journal}{Phys. Rev.} \textbf{\bibinfo{volume}{D96}},
  \bibinfo{pages}{096025} (\bibinfo{year}{2017}), \eprint{1703.01991}.

\bibitem[{\citenamefont{Aidala et~al.}(2017)}]{Aidala:2017pum}
\bibinfo{author}{\bibfnamefont{C.}~\bibnamefont{Aidala}} \bibnamefont{et~al.}
  (\bibinfo{collaboration}{PHENIX Collaboration}), \bibinfo{journal}{Phys. Rev.}
  \textbf{\bibinfo{volume}{D95}}, \bibinfo{pages}{112001}
  (\bibinfo{year}{2017}), \eprint{1703.09333}.

\bibitem[{\citenamefont{Aidala et~al.}(2018)}]{Aidala:2018gmp}
\bibinfo{author}{\bibfnamefont{C.}~\bibnamefont{Aidala}} \bibnamefont{et~al.}
  (\bibinfo{collaboration}{PHENIX Collaboration}), \bibinfo{journal}{Phys. Rev.}
  \textbf{\bibinfo{volume}{D98}}, \bibinfo{pages}{012006}
  (\bibinfo{year}{2018}), \eprint{1805.01491}.

\bibitem[{\citenamefont{Collins}(1993)}]{Collins:1992kk}
\bibinfo{author}{\bibfnamefont{J.~C.} \bibnamefont{Collins}},
  \bibinfo{journal}{Nucl. Phys.} \textbf{\bibinfo{volume}{B396}},
  \bibinfo{pages}{161} (\bibinfo{year}{1993}).

\bibitem[{\citenamefont{Bacchetta et~al.}(2004)\citenamefont{Bacchetta,
  D'Alesio, Diehl, and Miller}}]{Bacchetta:2004jz}
\bibinfo{author}{\bibfnamefont{A.}~\bibnamefont{Bacchetta}},
  \bibinfo{author}{\bibfnamefont{U.}~\bibnamefont{D'Alesio}},
  \bibinfo{author}{\bibfnamefont{M.}~\bibnamefont{Diehl}}, \bibnamefont{and}
  \bibinfo{author}{\bibfnamefont{C.~A.} \bibnamefont{Miller}},
  \bibinfo{journal}{Phys. Rev.} \textbf{\bibinfo{volume}{D70}},
  \bibinfo{pages}{117504} (\bibinfo{year}{2004}), \eprint{hep-ph/0410050}.

\bibitem[{\citenamefont{Anselmino
  et~al.}(2006{\natexlab{b}})\citenamefont{Anselmino, D'Alesio, Melis, and
  Murgia}}]{Anselmino:2006yq}
\bibinfo{author}{\bibfnamefont{M.}~\bibnamefont{Anselmino}},
  \bibinfo{author}{\bibfnamefont{U.}~\bibnamefont{D'Alesio}},
  \bibinfo{author}{\bibfnamefont{S.}~\bibnamefont{Melis}}, \bibnamefont{and}
  \bibinfo{author}{\bibfnamefont{F.}~\bibnamefont{Murgia}},
  \bibinfo{journal}{Phys. Rev.} \textbf{\bibinfo{volume}{D74}},
  \bibinfo{pages}{094011} (\bibinfo{year}{2006}{\natexlab{b}}),
  \eprint{hep-ph/0608211}.

\bibitem[{\citenamefont{Anselmino
  et~al.}(2005{\natexlab{b}})\citenamefont{Anselmino, Boglione, D'Alesio, Kotzinian, Murgia, and Prokudin}}]{Anselmino:2005nn}
\bibinfo{author}{\bibfnamefont{M.}~\bibnamefont{Anselmino}},
 \bibinfo{author}{\bibfnamefont{M.}~\bibnamefont{Boglione}},
  \bibinfo{author}{\bibfnamefont{U.}~\bibnamefont{D'Alesio}},
  \bibinfo{author}{\bibfnamefont{A.}~\bibnamefont{Kotzinian}},
  \bibinfo{author}{\bibfnamefont{F.}~\bibnamefont{Murgia}}, \bibnamefont{and}
  \bibinfo{author}{\bibfnamefont{A.}~\bibnamefont{Prokudin}},
  \bibinfo{journal}{Phys. Rev.}
  \textbf{\bibinfo{volume}{D71}}, \bibinfo{pages}{074006}
  (\bibinfo{year}{2005}{\natexlab{b}}), \eprint{hep-ph/0501196}.

\bibitem[{\citenamefont{Pumplin et~al.}(2002)\citenamefont{Pumplin, Stump,
  Huston, Lai, Nadolsky, and Tung}}]{Pumplin:2002vw}
\bibinfo{author}{\bibfnamefont{J.}~\bibnamefont{Pumplin}},
  \bibinfo{author}{\bibfnamefont{D.~R.} \bibnamefont{Stump}},
  \bibinfo{author}{\bibfnamefont{J.}~\bibnamefont{Huston}},
  \bibinfo{author}{\bibfnamefont{H.~L.} \bibnamefont{Lai}},
  \bibinfo{author}{\bibfnamefont{P.~M.} \bibnamefont{Nadolsky}},
  \bibnamefont{and} \bibinfo{author}{\bibfnamefont{W.~K.} \bibnamefont{Tung}},
  \bibinfo{journal}{JHEP} \textbf{\bibinfo{volume}{07}}, \bibinfo{pages}{012}
  (\bibinfo{year}{2002}), \eprint{hep-ph/0201195}.

\bibitem[{\citenamefont{de~Florian et~al.}(2007)\citenamefont{de~Florian,
  Sassot, and Stratmann}}]{deFlorian:2007aj}
\bibinfo{author}{\bibfnamefont{D.}~\bibnamefont{de~Florian}},
  \bibinfo{author}{\bibfnamefont{R.}~\bibnamefont{Sassot}}, \bibnamefont{and}
  \bibinfo{author}{\bibfnamefont{M.}~\bibnamefont{Stratmann}},
  \bibinfo{journal}{Phys. Rev.} \textbf{\bibinfo{volume}{D75}},
  \bibinfo{pages}{114010} (\bibinfo{year}{2007}), \eprint{hep-ph/0703242}.

\bibitem[{\citenamefont{D'Alesio
  et~al.}(2011{\natexlab{b}})\citenamefont{D'Alesio, Murgia, and
  Pisano}}]{DAlesio:2010sag}
\bibinfo{author}{\bibfnamefont{U.}~\bibnamefont{D'Alesio}},
  \bibinfo{author}{\bibfnamefont{F.}~\bibnamefont{Murgia}}, \bibnamefont{and}
  \bibinfo{author}{\bibfnamefont{C.}~\bibnamefont{Pisano}},
  \bibinfo{journal}{Phys. Rev.} \textbf{\bibinfo{volume}{D83}},
  \bibinfo{pages}{034021} (\bibinfo{year}{2011}{\natexlab{b}}),
  \eprint{1011.2692}.

\bibitem[{\citenamefont{Koike and Yoshida}(2011)}]{Koike:2011mb}
\bibinfo{author}{\bibfnamefont{Y.}~\bibnamefont{Koike}} \bibnamefont{and}
  \bibinfo{author}{\bibfnamefont{S.}~\bibnamefont{Yoshida}},
  \bibinfo{journal}{Phys. Rev.} \textbf{\bibinfo{volume}{D84}},
  \bibinfo{pages}{014026} (\bibinfo{year}{2011}), \eprint{1104.3943}.

\bibitem[{\citenamefont{Brodsky et~al.}(2013)\citenamefont{Brodsky, Fleuret,
  Hadjidakis, and Lansberg}}]{Brodsky:2012vg}
\bibinfo{author}{\bibfnamefont{S.~J.} \bibnamefont{Brodsky}},
  \bibinfo{author}{\bibfnamefont{F.}~\bibnamefont{Fleuret}},
  \bibinfo{author}{\bibfnamefont{C.}~\bibnamefont{Hadjidakis}},
  \bibnamefont{and} \bibinfo{author}{\bibfnamefont{J.~P.}
  \bibnamefont{Lansberg}}, \bibinfo{journal}{Phys. Rept.}
  \textbf{\bibinfo{volume}{522}}, \bibinfo{pages}{239} (\bibinfo{year}{2013}),
  \eprint{1202.6585}.

\bibitem[{\citenamefont{Hadjidakis et~al.}(2018)}]{Hadjidakis:2018ifr}
\bibinfo{author}{\bibfnamefont{C.}~\bibnamefont{Hadjidakis}} \bibnamefont{et~al.}
  (\bibinfo{year}{2018}), \eprint{1807.00603}.

\bibitem[{\citenamefont{Di~Nezza}(2017)}]{LHCbspin:2017xx}
\bibinfo{author}{\bibfnamefont{P.}~\bibnamefont{Di~Nezza}},
  \bibinfo{journal}{https://indico.cern.ch/event/644287/contributions/2747717}
  (\bibinfo{year}{2017}).

\bibitem[{\citenamefont{Di~Nezza}(2018)}]{LHCbspin:2018xx}
\bibinfo{author}{\bibfnamefont{P.}~\bibnamefont{Di~Nezza}},
  \bibinfo{journal}{Proceedings of the XXIII International Spin Symposium (SPIN
  2018), 9-14 Sept. 2018, Ferrara (Italy)}  (\bibinfo{year}{2018}).

\bibitem[{\citenamefont{Schmidt et~al.}(2005)\citenamefont{Schmidt, Soffer, and
  Yang}}]{Schmidt:2005gv}
\bibinfo{author}{\bibfnamefont{I.}~\bibnamefont{Schmidt}},
  \bibinfo{author}{\bibfnamefont{J.}~\bibnamefont{Soffer}}, \bibnamefont{and}
  \bibinfo{author}{\bibfnamefont{J.-J.} \bibnamefont{Yang}},
  \bibinfo{journal}{Phys. Lett.} \textbf{\bibinfo{volume}{B612}},
  \bibinfo{pages}{258} (\bibinfo{year}{2005}), \eprint{hep-ph/0503127}.

\bibitem[{\citenamefont{Godbole et~al.}(2018)\citenamefont{Godbole, Kaushik,
  Misra, and Padval}}]{Godbole:2018mmh}
\bibinfo{author}{\bibfnamefont{R.~M.} \bibnamefont{Godbole}},
  \bibinfo{author}{\bibfnamefont{A.}~\bibnamefont{Kaushik}},
  \bibinfo{author}{\bibfnamefont{A.}~\bibnamefont{Misra}}, \bibnamefont{and}
  \bibinfo{author}{\bibfnamefont{S.}~\bibnamefont{Padval}}
  (\bibinfo{year}{2018}), \eprint{1810.07113}.

\bibitem[{\citenamefont{Bomhof}(2007)}]{Bomhof:2007zz}
\bibinfo{author}{\bibfnamefont{C.~J.} \bibnamefont{Bomhof}}, Ph.D. thesis,
  \bibinfo{school}{Vrije U., Amsterdam} (\bibinfo{year}{2007}).

\end{thebibliography}

\end{document}